\documentclass[]{aa}
\usepackage{txfonts}
\usepackage{natbib}
\usepackage{graphicx}
\usepackage{psfig}
\usepackage{amssymb}

\def\src{IGR~J08408-4503}

\def\xmm{{\em XMM--Newton}}

\def\inte{{\em INTEGRAL}}
\def\suz{{\em Suzaku}}

\def\gaia{{\em Gaia}}
\def\epic{{EPIC}}
\def\pn{{pn}}
\def\mos{{MOS}}
\def\mosuno{{MOS1}}
\def\mosdue{{MOS2}}

\def\approxgt{\mathrel{\hbox{\rlap{\lower.55ex \hbox {$\sim$}}
        \kern-.3em \raise.4ex \hbox{$>$}}}}
\def\approxlt{\mathrel{\hbox{\rlap{\lower.55ex \hbox {$\sim$}}
        \kern-.3em \raise.4ex \hbox{$<$}}}}

\def\flux {\mbox{erg cm$^{-2}$ s$^{-1}$}}
\def\lum {\mbox{erg s$^{-1}$}}

\def\ltsima{$\; \buildrel < \over \sim \;$}
\def\lsim{\lower.5ex\hbox{\ltsima}}
\def\gtsima{$\; \buildrel > \over \sim \;$}
\def\gsim{\lower.5ex\hbox{\gtsima}}

\def\hcm {\hbox {\ifmmode $ atom cm$^{-2}\else atom cm$^{-2}$\fi}}

\def\arcsec {\hbox{$^{\prime\prime}$}}

\def \apj {ApJ}
\def \aj {AJ}
\def \apjl {ApJL}
\def \apjs {ApJS}
\def \aap {A\&A}

\def \mnras {MNRAS}

\def \ssr {Space Science Reviews}

\newcommand{\be}{\begin{equation}}

\newcommand{\ee}{\end{equation}}

\newcommand{\swift}{{\emph{Swift}}}

\begin{document}
%%%%%%%%%%%%%%%%%%%%%%%%%%%%%%%%%%%%%%%%%%%%%%%%
\title{Detecting the intrinsic X-ray emission from the O-type donor star and
the residual accretion in a Supergiant Fast X-ray Transient during its faintest state
\thanks{
Based on observations (ObsID 0861460101) obtained with {\em XMM-Newton}, an 
ESA science mission with instruments and contributions directly funded by ESA Member States and NASA.
}}

\author{L.~Sidoli\inst{1},   K.~Postnov\inst{2,3}, L.~Oskinova\inst{4,3}, P. Esposito\inst{5,1}, A. De Luca\inst{1,6}, M. Marelli\inst{1} and R. Salvaterra\inst{1} 
}
\institute{$^1$ INAF, Istituto di Astrofisica Spaziale e Fisica Cosmica, via A.\ Corti 12, I-20133 Milano, Italy
\\
$^2$ Sternberg Astronomical Institute, Lomonosov Moscow State University, Universitetskij pr. 13, 119234 Moscow, Russia\\
$^3$ Kazan Federal University, Kremlyovskaya 18, 420008 Kazan, Russia \\
$^4$ Institute for Physics and Astronomy, University Potsdam, 14476 Potsdam, Germany \\
$^5$ Scuola Universitaria Superiore IUSS Pavia, Piazza della Vittoria 15, 27100, Pavia, Italy \\
$^6$ INFN, Sezione di Pavia, via A. Bassi 6, I-27100 Pavia, Italy
}

\offprints{L. Sidoli, lara.sidoli@inaf.it}

\date{Received 25 May 2021/ Accepted 22 June 2021}

\authorrunning{L. Sidoli et al.}

\titlerunning{\xmm\ observes \src\ in quiescence}

\abstract{
We report on the results of an \xmm\ observation of the Supergiant Fast X-ray Transient (SFXT) \src\ performed  in June 2020. 
The source is composed by a compact object (likely a neutron star) orbiting around an O8.5Ib-II(f)p star, LM Vel.
The X-ray light curve shows a very low level of emission, punctuated by a single, faint flare. 
Analysis of spectra measured during the flare and during quiescence is performed.
The quiescent state shows a continuum spectrum well deconvolved to  three spectral models: two components are from a collisionally-ionized plasma 
(with temperatures kT$_1$=0.24 keV and kT$_2$=0.76 keV),
together with a power law model (photon index, $\Gamma$, of $\sim$2.55), dominating above $\sim$2 keV. 
The X-ray flux emitted at this lowest level is 3.2$\times10^{-13}$ \flux (0.5-10 keV, corrected for the interstellar absorption),
implying an X-ray luminosity of 1.85$\times10^{32}$ \lum\ (at 2.2 kpc).
The two temperature collisionally-ionized plasma is intrinsic to the stellar wind of the donor star, while the power law can be interpreted as emission due to residual, low level accretion onto the compact object.
The X-ray luminosity contributed by the power law component only, 
in the lowest state, is (4.8$\pm{1.4})\times10^{31}$ \lum,
the lowest quiescent luminosity detected from the compact object in an SFXT.
Thanks to this very faint X-ray state caught by \xmm, 
X-ray emission from the wind of the donor star LM Vel could be well-established 
and studied in detail for the first time, as well as a very low level of accretion onto the compact object.
The residual accretion rate onto the compact object in \src\ can be interpreted as the Bohm diffusion of (possibly magnetized) plasma entering the neutron star magnetosphere at low Bondi capture rates from the supergiant donor wind at the quasi-spherical radiation-driven settling accretion stage.
\keywords{stars: neutron - X-rays: binaries - pulsars: individual: \src, HD 74194, LM Vel}
}

\maketitle

        %%%%%%%%%%%%%%%%%%%%%%%%%%%%%%%%%%%%%%%%%%%%%%%%%%%%%%%%%
        \section{Introduction\label{intro}}
        %%%%%%%%%%%%%%%%%%%%%%%%%%%%%%%%%%%%%%%%%%%%%%%%%%%%%%%%%

\src\ is a transient X-ray source discovered during outburst in 2006 thanks to observations performed by the \inte\ satellite  \citep{Goetz2006}. 
Not long after, the analysis of \inte\ archival data revealed an earlier
outburst that occurred  in 2003  \citep{Mereghetti2006}, and thus unveiled a 
recurrent bright, flaring X-ray activity of the source.
\src\ was associated with the O-type supergiant LM Vel (also known as HD 74194; 
\citealt{Goetz2006, Masetti2006, Barba2006, Kennea2006}) 
leading  to its classification as a Supergiant Fast X-ray Transient
(SFXT), i.e.\ a new type of  high mass X-ray binaries (HMXBs; \citealt{2019NewAR..8601546K}) discovered during the  \inte\ monitoring of the Galactic plane \citep{Sguera2005, Sguera2006, Negueruela2005a}. 

SFXTs are detected by \inte\ during bright flares lasting a quite short interval of time ($\sim$1-2 ks) and, usually, reaching  peak luminosities 
of 10$^{36}$-10$^{37}$ \lum\ (see \citealt{Sidoli2017review, Sidoli2018} for reviews). 
The SFXT flares can be part of rare (less than 5\% of the time, \citealt{Sidoli2018}) longer outbursts, with a duration of a few days \citep{Romano2007, Rampy2009:suzaku17544, Sidoli2016b}. 

The monitoring campaigns with the Neil Gehrels \swift\ satellite have shown that the most frequent X-ray state in SFXTs is at  luminosities below a few $\times 10^{34}$ \lum\ \citep{2008ApJ...687.1230S, 2015JHEAp...7..126R},  down to X-ray luminosities of $\sim$10$^{32}$ \lum\ in some members of the class \citep{zand2005, Bozzo2010, Sidoli2010igr08408}. 
This huge range of X-ray variability is a characterizing property of the SFXT 
class (an updated list of the dynamic range of X-ray fluxes and duty cycles in 
SFXTs, compared with other types of HMXBs can be found in \citealt{Sidoli2018}).
Nowadays, the SFXT class has about twenty confirmed members, plus a similar 
number of candidates still missing optical or infrared (IR) 
identifications.

In persistent HMXBs with supergiant massive donor stars,  
the high X-ray luminosity is sustained by accretion of stellar wind material onto the compact object, usually a neutron star (NS). 
The massive star winds are quite stationary and uniform when averaged over years \citep{Lamers1999}. However, on shorter time scales (days to
hours), the non-stationary processes, such as shocks, are operating in
stellar winds \citep{Feldmeier1997}. Moreover, the large scale
structures corotating with the star as well as the small scale
inhomogeneities (often referred to as clumps) are ubiquitous in
radiatively driven winds \citep{2008A&ARv..16..209P, 2019ApJ...873...81M, 2021MNRAS.504.2051V}.

SFXTs are HMXBs with a compact object orbiting an OB supergiant companion, but behave in a completely different way.
The physical mechanism responsible for the SFXT phenomenology continues to be debated:  the quasi-spherical settling accretion regime \citep{2012MNRAS.420..216S, 2014MNRAS.442.2325S}, the propeller mechanism and the magnetic barrier \citep{Grebenev2007, Bozzo2008} are the most discussed theoretical explanations, to date. These mechanisms are able to reduce the accretion of wind material onto 
the NS, for most of the time. 
Nevertheless, the issue remains controversial because in almost all SFXTs the  important properties of the donor star and of the compact object are unknown.

So far, only a few systems have been spectroscopically observed at optical, 
IR and ultra-violet (UV) wavelengths  and analyzed using modern stellar atmosphere models required to determine characteristics of the donor's wind \citep[e.g.][]{2016A&A...591A..26G, Hainich2020}. 
These studies showed that the winds of donor stars do not drastically differ from their single star counterparts -- the winds are clumped, and have usual mass-loss rates and wind velocities. However, the radiation from accreting compact object located near the donor star has a noticeable effect on outer stellar atmosphere 
and can change the wind ionization and acceleration
\citep{Sander2019}. 
All early type OB supergiants stars are intrinsic X-ray sources 
\citep{Berg1997} emitting soft X-rays at $L_{\rm X} \sim 10^{32}$\,\lum.  
Therefore, it might be expected that the intrinsic X-ray emission from stellar 
winds could be detected also in HMXBs, if X-rays
produced by accretion do not significantly outshine radiation from the donor's wind.  

The orbital geometry in SFXTs remains unclear: while orbital periods have been measured in about a half of the systems (from the periodic modulation of their X-ray long-term light curve),  the eccentricity is largely unknown (see the reviews by \citealt{Walter2015}, \citealt{2017SSRv..212...59M}, \citealt{Sidoli2018} and \citealt{2019NewAR..8601546K} for updated lists of their orbital periods).

Among all SFXTs, the subject of this study, \src, provides one of the most important laboratories for investigating the physics governing SFXT phenomenology. It is the only SFXT with a well determined orbital geometry: its orbit is very eccentric, $e=0.63\pm{0.03}$, with $9.5436\pm{0.0002}$\,d orbital period \citep{Gamen2015}. 
The distance is also well known: according to \gaia\ Early Data Release 3 \citep{2021A&A...649A...1G} 
it is located at a distance of $2.20^{+0.08}_{-0.09}$\,kpc  \citep{2021AJ....161..147B}.

\src\ has an extremely low duty cycle. 
\inte\ data show that it undergoes bright (i.e. above a flux of $\sim$3$\times10^{-10}$ \flux\ in the energy range 18-50\,keV) X-ray flares  for only 0.09\%\ of the time \citep{Sidoli2018}. 
The outbursts have occurred in a broad range of orbital phases ($\Delta\phi \approx \pm{0.15}$) around periastron (a collection of bright X-ray flares is reported by \citealt{Gamen2015} and \citealt{Ducci2019}). 
Nevertheless, it is important to remark that the passage at periastron does not 
necessarily trigger an outburst: this is testified by a very long observation performed with \suz\ around periastron when the X-ray luminosity remained in the range 10$^{32}$-10$^{33}$ \lum\  \citep{Sidoli2010igr08408}.
\src\ spends about 68\% of the time having X-ray luminosity below
1.1$\times$10$^{33}$ \lum\ in the 2-10 keV energy band \citep{2015JHEAp...7..126R}. 
X-ray pulsations have not been detected in \src, therefore  the nature 
of the compact object is not firmly established.  Nevertheless, a NS is usually assumed (as in other SFXTs; \citealt{Sidoli2017review}), since the X-ray spectrum in outburst resembles the spectra  of accreting pulsars in HMXBs \citep{Goetz2007, Sidoli2009_swift, Romano2009}.

As for the whole SFXT class, the physical mechanism driving the \src\ behavior is heavily debated:  the previous literature  has 
investigated the supergiant clumpy winds  \citep{Hainich2020},  
the role played by the eccentric orbit \citep{2021MNRAS.501.2403B}, 
and  the possibility  of an accretion disc formation 
fed by Roche lobe overflow \citep{Ducci2019}.

In this paper we report the \xmm\ observation of \src\ performed in June 2020 (Sect.\,\ref{data}), 
around the orbital phase $\sim$0.65 just after the apastron passage, 
catching this SFXT in a very low luminosity state. 
 The timing and spectral analysis are reported in Sect.\,\ref{timing} and \ref{sec:spec}, while in Sect.\,\ref{sec:quiesc} we interpret the lowest X-ray emission in terms of a twofold contribution: X-rays from the donor wind and residual accretion onto the compact object, the lowest luminosity state ever detected from a SFXT. The comparison with previous observations of faint X-ray emission in \src\ is outlined in Sect.\,\ref{Sect:comp}. The physical picture explaining the very low accretion rate observed by \xmm\ is discussed in Sects.\,\ref{Sect:physics} and \ref{Sect:applied}. Finally, Sect.\,\ref{Sect:concl} presents our conclusions.

         %%%%%%%%%%%%%%%%%%%%%%%%%%%%%%%%%%%%%%%%%%%%%%%%%%%%%%%%%%%%%%%%%%%%
         \section{Observation and data reduction}
         \label{data}
         %%%%%%%%%%%%%%%%%%%%%%%%%%%%%%%%%%%%%%%%%%%%%%%%%%%%%%%%%%%%%%%%%%%%

\src\ was observed by \xmm\ \citep{Jansen2001} in June 2020, with a net exposure time of $\sim$47\,ks
(details of the \epic\ exposures are reported in Table~\ref{tab:log}).
All \epic\ cameras \citep{Struder2001, Turner2001} operated in full frame mode and using the medium filter.

Data (Obs.ID 0861460101) were reprocessed using the version 18 
of the \xmm\ Science Analysis Software (SAS), adopting
standard procedures.
The response and ancillary matrices
were generated with {\em rmfgen} and {\em arfgen} available in the SAS.
A part of the exposure time at the start 
of the observation which was affected by the high background  
was excluded from the extraction of all source products. 
 
Source light curves and spectra were extracted 
from circular regions centered on the source emission,
with a 30\arcsec\ radius for the \pn\ and 
60\arcsec\ for the two \mos, 
and selecting pattern from 0 to 4 (\epic\ \pn), and from 0 to 12 (\mos).
Background spectra were obtained from similar size regions offset from the source position.
\src\ was observed at a very low state (in Table~\ref{tab:log}, last column,  the time-averaged \epic\ 
net count rates are listed), too faint for a meaningful spectroscopy with the Reflection Grating Spectrometer \citep{DenHerder2001}.

\epic\ spectra were simultaneously fitted in the energy range 0.3-12 keV using {\sc xspec} (version 12.10.1; \citealt{Arnaud1996}), 
including cross-calibration constants to take into account calibration uncertainties. 
All fluxes were estimated in the 0.5-10 keV range, for consistency with previous literature.
When fitting the spectra, the absorption model {\sc TBabs} was adopted, assuming the photoelectric absorption cross sections of \cite{Verner1996} and the interstellar abundances of \cite{Wilms2000}. 
The spectra were rebinned to have at least 25 counts per bin, to apply the $\chi^{2}$ statistics. 
All  uncertainties  in the spectral analysis are given 
at 90\% confidence level, for one interesting parameter.
The uncertainty on the unabsorbed X-ray fluxes have been obtained using {\sc cflux} in {\sc xspec}.

All luminosities have been calculated assuming a source distance of 2.2 kpc.

%%%%%%%%%% ----------------------------------------------------------------
\begin{table*}
\caption[]{Summary of the \xmm\ observation targeted on \src. 
}
\begin{tabular}{lllll}
\hline
\noalign {\smallskip}
Instrument       &  Exposure Start Time  (UTC)       & Stop time (UTC) 	             & Exposure     & Net rates (0.3-12 keV) \\
		 &   (yyyy-mm-dd hh:mm:ss)    	     &  (yyyy-mm-dd hh:mm:ss)        & (ks)         & (counts s$^{-1}$)   \\      
\hline
\noalign {\smallskip}
 \epic\ \pn           &     2020-06-01 20:12:10	&   2020-06-02 08:38:13          &	44.8	    & (8.63$\pm{0.24}$)$\times10^{-2}$  \\
 \epic\ \mosuno       &     2020-06-01 19:46:28 &   2020-06-02 08:43:36          &      46.6 	    & (2.27$\pm{0.11}$)$\times10^{-2}$  \\
 \epic\ \mosdue       &     2020-06-01 19:46:49 &   2020-06-02 08:43:39          &      46.6	    & (2.21$\pm{0.11}$)$\times10^{-2}$  \\
%------------------------------------------------------------
\hline
\label{tab:log}
\end{tabular}
\end{table*}
%%%%%%%%%% -------------------------------------------------

%%%%%%%%%%%%%%%%%%%%%%%%%%%%%%%%%%%%%%%%%%%%%%%%%%%%%%%%%%%%%%%%%%%%
\section{Temporal analysis}
\label{timing}
%%%%%%%%%%%%%%%%%%%%%%%%%%%%%%%%%%%%%%%%%%%%%%%%%%%%%%%%%%%%%%%%%%%%

\begin{figure}
\begin{center}
\includegraphics[width=5.75cm,angle=-90]{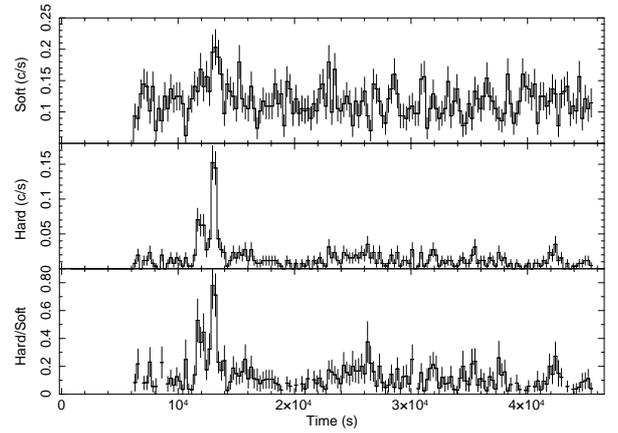}
\caption{Source light curve  (\epic\ \pn, not background subtracted, bin time=256 s) 
in two energy ranges (Soft=0.3-2 keV, Hard=2-12 keV), together with their hardness ratio.
The first $\sim$6000 s of the observation have been excluded, because of high background level. 
}
\label{fig:2lc_ratio}
\end{center}
\end{figure}

The source was caught by \xmm\ in a very low emission state. 
The light curve in two energy ranges (above and below 2 keV) is shown in Fig.~\ref{fig:2lc_ratio}, together with their hardness ratio.
A single faint source flare was observed, after about 12 ks from the beginning of the \epic\ \pn\ exposure.
Harder emission is evident during the flare.

We searched the data for the presence of periodic modulations by means of a Fourier transform. Since we ascertained that both the source flare around 12\,ks and the data gap produced by its removal cause strong non-Poissonian noise in the power density spectrum, we used only the part of the observation after the flare, for an exposure of about 29.3\,ks.  
We did not find any candidate signal and set for a sinusoidal modulation an upper limit on the pulsed fraction (defined as the semi-amplitude of the sinusoidal profile divided by the mean count rate) of $\approx$30\% for periods from $\sim$0.15 to 15\,000\,s (in the 1--10\,keV energy range; we used only the \epic\ \pn\ data below 5.4\,s and the combined \pn\ and \mos\ data above).
No other structures, such as quasi-periodic oscillation features, are discernible in the same range in the Fourier power density spectra.

%%%%%%%%%%%%%%%%%%%%%%%%%%%%%%%%%%%%%%%%%%%%%%%%%%%%%
%%%%%%%%%%%%%%%%%%%%%%%%%%%%%%%%%%%%%%%%%%%%%%%%%%%%%% 

%%%%%%%%%%%%%%%%%%%%%%%%%%%%%%%%%%%%%%%%%%%%%%%%%%%%%
\begin{figure}
\begin{center}
\includegraphics[width=5.75cm,angle=-90]{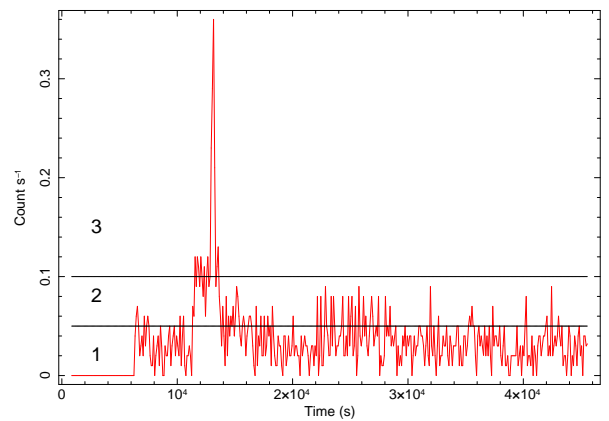}
\caption{Source light curve  
(\epic\ \pn, 1-10 keV, not background subtracted, bin time=100 s), 
where the horizontal lines mark
the three count rate intervals adopted for the intensity selected spectroscopy:
interval 1 (lowest state), 2 (intermediate state) and 3 (the faint flare).
}
\label{fig:lc4intselspec}
\end{center}
\end{figure}
%%%%%%%%%%%%%%%%%%%%%%%%%%%%%%%%%%%%%%%%%%%%%%%%%%%%%% 

    %%%%%%%%%%%%%%%%%%%%%%%%%%%%%%%%%%%%%%%%%%%%%%%%%%%%%
    \section{Spectroscopy}
    \label{sec:spec}
    %%%%%%%%%%%%%%%%%%%%%%%%%%%%%%%%%%%%%%%%%%%%%%%%%%%%%
 
During the flare event, the source hardness ratio has increased  (Fig.~\ref{fig:2lc_ratio}, bottom panel) prompting us to perform spectroscopic analyses at three different states of the source flux 
selected using the  \epic\ \pn\ light curve (Fig.~\ref{fig:lc4intselspec}).
We extracted spectra from the following intervals: below 0.05\,count s$^{-1}$ (lowest state, Sect.~\ref{sec:quiesc}),
between 0.05 and 0.1\,count s$^{-1}$ (intermediate state, Sect.~\ref{sec:interspec}), and above 0.1\,count s$^{-1}$ (Sect.~\ref{sec:flarespec}).

    	%%%%%%%%%%%%%%%%%%%%%%%%%%%%%%%%%%%%%%%%%%%%
    	\subsection{The lowest state \label{sec:quiesc}}
    	%%%%%%%%%%%%%%%%%%%%%%%%%%%%%%%%%%%%%%%%%%%%

\begin{figure}
\begin{center}
\includegraphics[width=5.75cm,angle=-90]{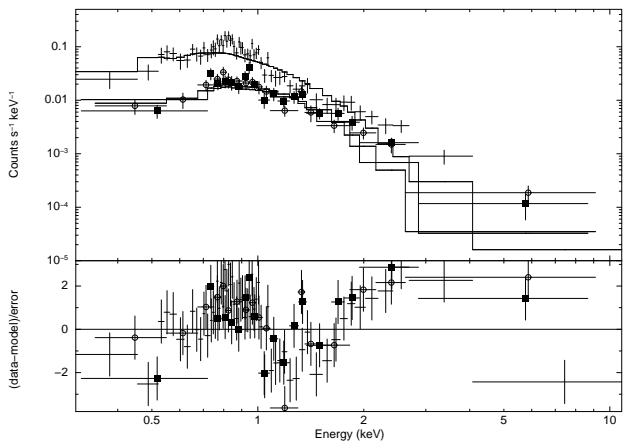}
\caption{\epic\ spectra extracted from the lowest state, fitted by a single 
absorbed power law model. In the upper panel we show the counts spectra, while in the lower panel the residuals 
 with respect to the model, in terms of standard deviation.
The meaning of the symbols is the following: crosses, empty circles and solid squares mark the EPIC pn, MOS1 and MOS2 spectra, respectively.
}
\label{fig:lowest_pow}
\end{center}
\end{figure}

The spectrum extracted from the data accumulated during the lowest state could not be fitted by any single spectroscopic model: for example, in Fig.~\ref{fig:lowest_pow} we show the residuals when a single, absorbed power law model is adopted.
Therefore, as a next step, we fitted the observed spectrum with various types of two-component models, each of them providing statistically good fits.  

The two-component models we consider are combinations of a thermal emission (a black body or a collisionally-ionized plasma model such as {\sc apec}), with  a second component, either a hotter thermal model or a power law. 

In the appendix, the spectral parameters obtained with six different models 
are shown in Table~\ref{tab:quiesc}.
We note that in a single case only (Model\,6 in Table~\ref{tab:quiesc}) two absorption components were needed to obtain an acceptable fit: the additional one was provided by a partial covering absorption model ({\sc pcfabs} in {\sc xspec}).

%%%%%%%%%%%%%%%%%%%%%%%%%%%%%%%%%%%%%%%%%%%%%%%%%%%%%
%%%%%%%%%%%%%%%%%%%%%%%%%%%%%%%%%%%%%%%%%%%%%%%%%%%%%% 

Since from a statistical point of view all six models provide  equally good descriptions of the spectrum at the lowest state, to choose the realistic 
model we need to take into account the nature and the properties of the source.
 
The donor star, LM Vel has the O8.5Ib-II(f)p spectral type \citep{Sota2014} and 
therefore, like other OB supergiants should be an intrinsic source of X-rays \citep{Harnden1979, Cassinelli1981}.
The commonly accepted explanation attributes X-rays to the radiatively driven winds of these stars. The wind instabilities lead to shocks and consequent heating of the part of 
wind matter to X-ray emitting temperatures \citep{Feldmeier1997}. The shocked wind plasma emits thermal X-rays. While this mechanism may not be as effective as previously thought \citep{Steinberg2018}, observationally, the X-ray properties of OB supergiants are well established \citep[e.g.][]{Berg1997, Naze2009, Oskinova2016}.  The typical X-ray spectrum of an O supergiant star is described by a multi-temperature optically thin plasma model, such as e.g.\ {\sc apec}. In case of the two-temperature models,  the temperature components $kT_1\approx 0.2$--$0.3$\,keV and $kT_2\approx 0.7$--$0.8$\,keV are typically found  \citep[e.g.][]{Rauw2015, Huenemoerder2020}.

\citet{Nebot2018} investigated the dependence of X-ray properties of O-stars on stellar and wind parameters. They show that the X-ray luminosity  of O-type supergiants is  
$\log{(L_{\rm X} [\lum])}\approx 32.7\pm 0.2$ 
and that it does not correlate with stellar bolometric luminosity. 

Thus, LM~Vel, is, by itself an intrinsic source of X-rays. 
Indeed, the well studied star $\zeta$~Ori has a similar spectral type, O9.7Ib. Its X-ray luminosity is 
$\log{(L_{\rm X} [\lum])}\approx 32.8$
and the X-ray spectrum is that of thermal coronal plasma with 
0.2\,keV and 0.8\,keV temperature components.  Another star with a similar spectral type, HD\,149404 (O8.5Iab) also has 
$\log{(L_{\rm X} [\lum])}\approx 32.7$.
Brief analysis of its archival {\em XMM-Newton} observations shows that its X-ray spectrum could be well described by a two-temperature {\sc apec} model with $\approx 0.2$\,keV and $\approx 0.8$\,keV components. 

In addition, somewhat hotter plasma components are, sometimes, measured in the spectra of
OB+OB binaries where stellar winds collide with each other \citep{Rauw2016}. However, to the best of our knowledge, non-thermal X-ray radiation  described by a power-law spectrum has not been observed neither in single nor binary OB supergiants with colliding winds.

None of the two-component models reported in Table\,\ref{tab:quiesc}  
is typical for an X-ray spectrum of a single OB supergiant. 
In particular, the two-temperature {\sc apec} model (Model 3) would be favoured, but it displays a plasma component with a temperature of $\sim$3 keV, which is much hotter than typically deduced from the X-ray spectroscopy of these stars.
Another issue with the two {\sc apec} models, is the resulting low absorbing column density, $N_{\rm H}$. The reddening towards LM Vel is well
established from the analysis of the UV spectra, $E(B-V)$=0.44 \citep{Hainich2020}.
Using $R_{\rm V}=3.1$, the extinction towards LM Vel is $A_{V}$=1.364\,mag.
The conversion between extinction and hydrogen equivalent absorbing column density has been investigated by several authors, sometimes with significantly different results 
(e.g.\ \citealt{Bohlin1978, Predehl1995, Vuong2003, Gudennavar2012, Liszt2014, Zhu2017, Foight2016}).
We use the relation found by \citet{Foight2016},
$N_{\rm H}$=2.87$\times10^{21}$\,$A_{V}$\,cm$^{-2}$\,mag$^{-1}$, 
to be  consistent with our overall spectral analysis, since these authors
adopt the interstellar abundances according to \citet{Wilms2000}. 
This implies a value $N_{\rm H}$=3.9$\times10^{21}$\,cm$^{-2}$ towards LM Vel.
Therefore, we fixed the absorbing column density to this value during the fitting with the two {\sc apec} models, but the resulting fit was poor, with positive residuals below 1 keV. 

\begin{figure}
\begin{center}
\includegraphics[width=5.8cm,angle=-90]{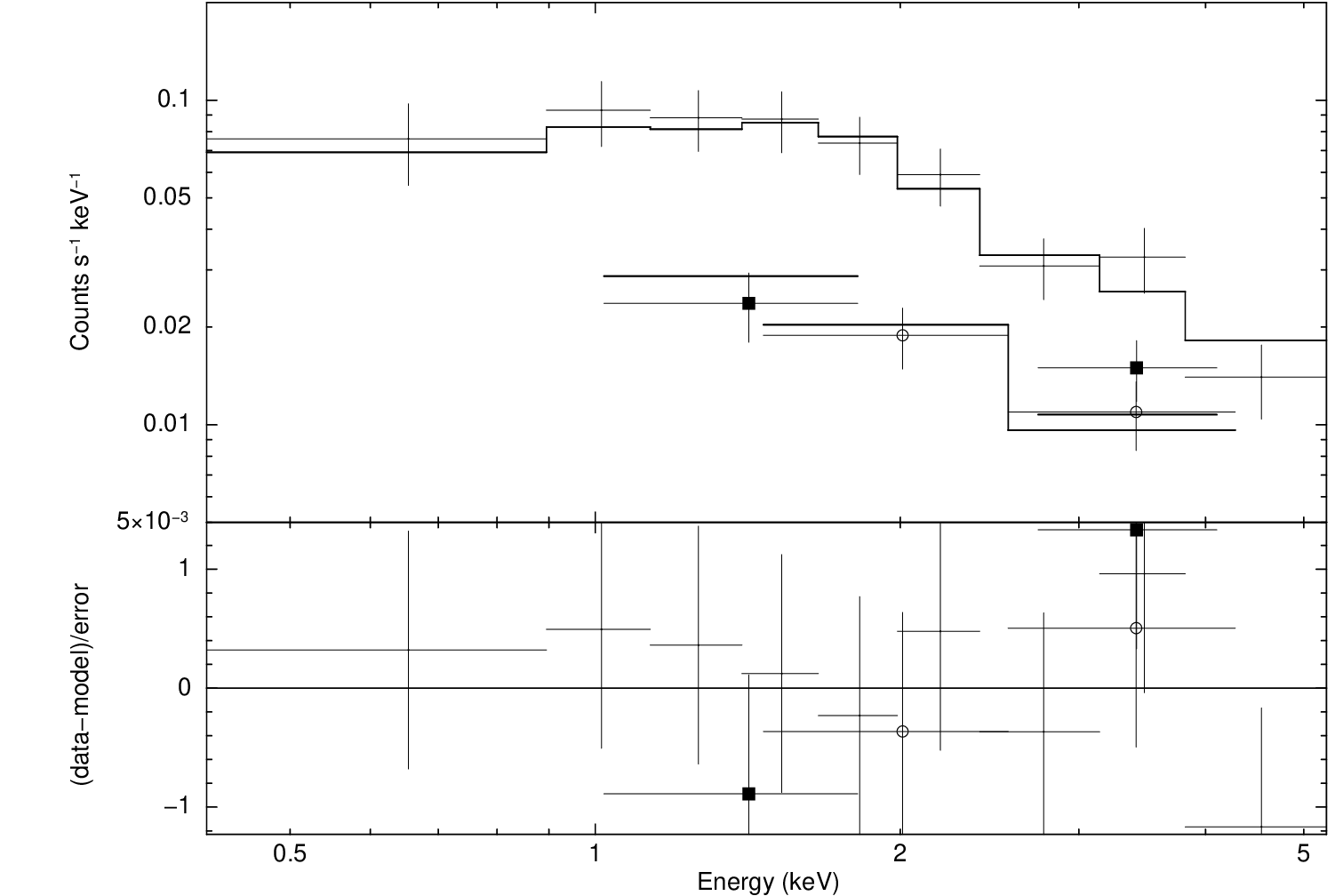} 
\includegraphics[width=5.8cm,angle=-90]{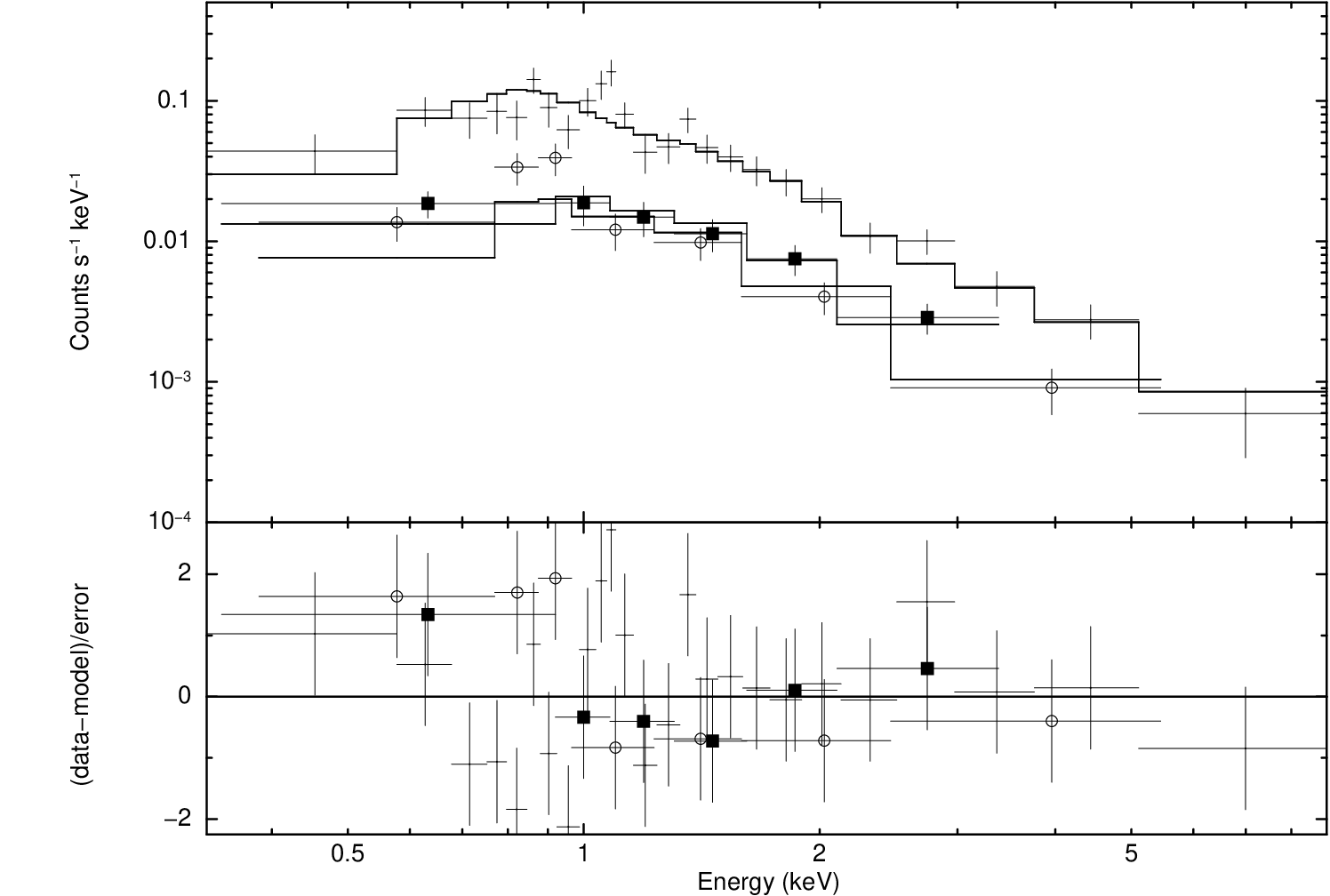} 
\includegraphics[width=5.8cm,angle=-90]{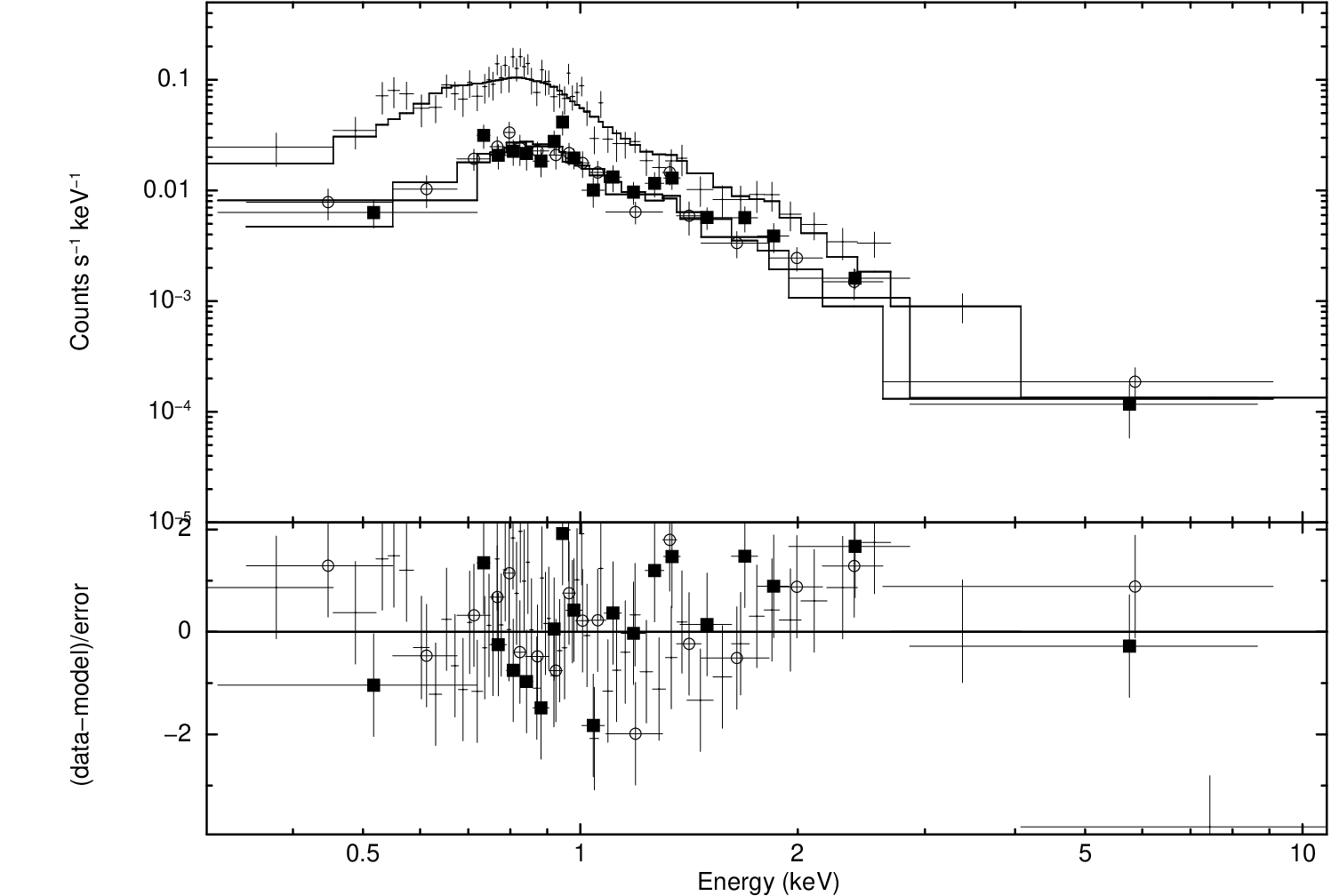}
     \caption{Best fit of the \src\ spectra extracted during the lowest luminosity state  (lower panel; 
     Model 2 in Table\,\ref{tab:quiesc_3apecs}), the intermediate state (middle panel; Model C in Table\,\ref{tab:interspec}) 
     and the faint flare (upper panel; Model B in Table\,\ref{tab:flarespec}).
The counts spectra are plotted together with the residuals in units of standard deviation.
The meaning of the symbols are the same as in Fig.~\ref{fig:lowest_pow}.}
\label{fig:bestfit_lowest}
\end{center}
\end{figure}

On the basis of these considerations, as the next step we consider three-component models.
We add a third continuum component to the two {\sc apec} models: either a third {\sc apec} model or a power law (see Table\,\ref{tab:quiesc_3apecs} for the spectral parameters resulting from these two models).
Since the double-temperature {\sc apec} plus the power law continuum resulted into a better description of the spectrum (Model\,2 in  Table\,\ref{tab:quiesc_3apecs}), this is our final model describing X-ray spectrum of \src\ during its lowest state (Fig.\,\ref{fig:bestfit_lowest}).

The physical interpretation of our favorite model is straightforward. 
The luminosity and the temperatures of the thermal {\sc apec} model components 
are consistent with the expectations from an intrinsic X-ray emission of the 
O-type supergiant donor star. We interpret the power law model component as emission produced by low-level  residual accretion onto the NS. 
The power law flux is  
$\sim$8$\times10^{-14}$ \flux (0.5-10 keV, corrected for the absorption),
implying a power law X-ray luminosity 
of (4.8$\pm{1.4})\times10^{31}$ \lum\ (this uncertainty 
accounts for the uncertainty on the flux only, not on the source distance).
 
We remark that, even fixing the absorbing column density to the lowest value towards LM Vel
($N_{\rm H}$=2.2$\times10^{21}$\,cm$^{-2}$, resulting from the relation 
$N_{\rm H}$=5$\times10^{21}$\,$E(B-V)$\,cm$^{-2}$\,mag$^{-1}$; \citealt{Vuong2003})
our conclusions do not change: the resulting temperatures of the two thermal components remain the same (within the uncertainties), as well as the photon index and power low
emitted flux. The only parameter  which would be significantly affected is the normalization of the softest thermal model: it would decrease to one third of the value reported in Table\,\ref{tab:quiesc_3apecs}, but
still consistent with X-ray emission from the supergiant wind.

%%%%%%%%%%%%%%%%%%%%%%%%%%%%%%%%%%%%%%%%%%%%%%%%%%%%%
%%%%%%%%%%%%%%%%%%%%%%%%%%%%%%%%%%%%%%%%%%%%%%%%%%%%%% 

%%%------------------------------------------------------------------
\begin{table*}
\caption{Spectroscopy of the  lowest luminosity state (\epic\ \pn, \mosuno\ and \mosdue),  fixing the absorption to the optical value
towards LM Vel (N$_{\rm H}$=3.9$\times$10$^{21}$ cm$^{-2}$).
}
\label{tab:quiesc_3apecs}
\vspace{0.0 cm}
\begin{center}
\begin{tabular}{lcc} \hline
 \hline
\noalign {\smallskip}
Param.                                   &  Model  1 $\,^{(d)}$                &          Model 2 $\,^{(d)}$       \\
\hline
\noalign {\smallskip}
N$_{\rm H}$ (10$^{22}$ cm$^{-2}$)          & $0.39$ (frozen)          &       $0.39$ (frozen)   \\
\multicolumn{2}{c}{----- APEC 1-------}   \\
kT$_{\rm APEC1}$   (keV)                  &  $0.24^{+0.04} _{-0.09}$    &    $0.24^{+0.04} _{-0.09}$      \\
norm$_{\rm APEC1}$ ($10^{-5}$ cm$^{-5}$)   &  $19.7^{+3.9} _{-3.9}$      &    $18.0^{+3.9} _{-4.0}$     \\
%----
EM$_{\rm APEC1}$$^a$ ($10^{54}$ cm$^{-3}$) &  $11.4^{+2.3} _{-2.3}$      &    $10.4^{+2.3} _{-2.3}$      \\
UF$_{\rm APEC1}$$^b$   (\flux)   & $1.8^{+0.4}_{-0.5}\times10^{-13}$     &    $1.7^{+0.4}_{-0.5}\times10^{-13}$      \\  
\multicolumn{2}{c}{----- APEC 2-------}   \\
kT$_{\rm APEC2}$   (keV)                  &  $0.75^{+0.15} _{-0.16}$     &    $0.76^{+0.17} _{-0.16}$       \\
norm$_{\rm APEC2}$ ($10^{-5}$ cm$^{-5}$)   &  $4.1^{+4.3} _{-2.2}$        &     $3.5^{+4.1} _{-2.1}$    \\
%----
EM$_{\rm APEC2}$$^a$ ($10^{54}$ cm$^{-3}$) &  $2.4^{+2.5} _{-1.2}$        &     $2.0^{+2.4} _{-1.2}$    \\
UF$_{\rm APEC2}$$^b$   (\flux)     & $8.1^{+7.7}_{-4.4}\times10^{-14}$    &   $6.6^{+1.1}_{-1.1}\times10^{-14}$      \\  
\multicolumn{2}{c}{----- APEC 3-------}   \\
kT$_{\rm APEC3}$   (keV)                   &  $2.8^{+1.3} _{-0.5}$   		& $-$	\\
norm$_{\rm APEC3}$ ($10^{-5}$ cm$^{-5}$)   &  $4.6^{+1.1} _{-1.0}$   		& $-$   \\
%----
EM$_{\rm APEC3}$$^a$ ($10^{54}$ cm$^{-3}$) &  $2.7^{+0.6} _{-0.7}$              & $-$	\\
UF$_{\rm APEC3}$$^b$   (\flux)     & $6.0^{+1.2}_{-1.2}\times10^{-14}$     & $-$	 \\  
\multicolumn{2}{c}{----- POWER LAW-------}   \\
$\Gamma$                             & $-$				 & 	 $2.55^{+0.38} _{-0.37}$ 	\\
norm$_{\rm pow}$  	        	     & $-$     				 &       $0.028^{+0.009} _{-0.008}$   	  \\
%----
UF$_{\rm pow}$$^b$   (\flux)         & $-$                               & $8.3^{+2.4}_{-2.0}\times10^{-14}$     \\  
L$_{\rm pow}$$^{a,b}$   (\lum)          &  $-$                               &  $4.8^{+1.4}_{-1.4}\times10^{31}$                          \\
\hline
UF$_{\rm total}$$^b$  (\flux)             & $3.21^{+0.62}_{-0.30}\times10^{-13}$  &   $3.20^{+0.30}_{-0.30}\times10^{-13}$      \\  
L$_{\rm total}$$^{a,b}$ (\lum)          & $1.85\times10^{32}$                   &  $1.85\times10^{32}$     \\   
\hline
UF$_{\rm apec1}$/ UF$_{\rm total}$        &   57\%                               &  54\%     \\
UF$_{\rm apec2}$/ UF$_{\rm total}$        &   25\%                               &  21\%      \\
UF$_{\rm pow}$/ UF$_{\rm total}$           &   $-$                               &  25\%      \\
\hline
$\chi^{2}_{\nu}$/dof                      &   1.020/142                        & 1.004/142	 \\
\hline
\hline
\end{tabular}
\footnotesize{\\
$^a$ A source distance of 2.2 kpc is assumed \\
$^b$ All fluxes and luminosities are in the energy range 0.5-10 keV and are corrected for the absorption \\
$^c$ Power law photon index. \\
$^d$ Two models are shown: Model 1 is made of three thermal plasma models  ({\sc  const * TBabs * (APEC1 + APEC2 + APEC3)}), Model 2 is composed of two thermal plasma models plus a power law ({\sc  const * TBabs * (APEC1 + APEC2 + PEGPWRLW)}).
}
\end{center}
\end{table*}
%%%%%%%%%% -------------------------------------------------

	%%%%%%%%%%%%%%%%%%%%%%%%%%%%%%%%%%%%%%%%%%%%
	\subsection{The intermediate state \label{sec:interspec}}
	%%%%%%%%%%%%%%%%%%%%%%%%%%%%%%%%%%%%%%%%%%%%

We have investigated the spectra extracted from the intermediate state (marked by
number 2 in Fig.~\ref{fig:lc4intselspec}), by adopting the same best fit continuum found in the spectroscopy
of the lowest X-ray emission: two thermal plasma models plus a power law (Model 2 in Table\,\ref{tab:quiesc_3apecs}). 
Since the two thermal models are consistent with emission from the wind of the companion star,
we fixed their parameters to the best fit found in the spectroscopy of the lowest luminosity state. 
The power law model parameters were allowed to vary during the fit, adopting three versions of this continuum model, as reported in 
Table~\ref{tab:interspec}:
in Model A, only the parameters of the power law are let free to vary; 
in Model B, also the absorbing column density of the overall continuum is allowed to vary (this because, sometimes, a larger
absorption can be  observed in HMXBs, than what is predicted from the optical extinction towards the donor star);
in Model C, an additional absorption of the power law model only is included, 
while the absorbing column density of the overall continuum is fixed to the value of LM Vel (Fig.\,\ref{fig:bestfit_lowest}).

The intermediate state can be explained with an increased flux from the power law component only, 
the contribution of which increases from 25\% of the total unabasorbed 0.5-10 keV radiation in the lowest state, to $\sim$60\%   in the intermediate state. 
The total X-ray luminosity reaches $\sim$3-4$\times10^{32}$ \lum\ during the intermediate state.

%%%-----------------------------------------------------------
\begin{table*}
  \caption{Spectroscopy of the intermediate luminosity state (\epic\ pn, \mosuno\ and \mosdue; Sect.~\ref{sec:interspec}).
    }
\label{tab:interspec}
\vspace{0.0 cm}
\begin{center}
\begin{tabular}{lccc} \hline
 \hline
\noalign {\smallskip}
Param.                             &  Model A               &          Model B                 &  Model C          \\
\hline
\noalign {\smallskip}
N$_{\rm H}$ (10$^{22}$ cm$^{-2}$)       & $0.39$ (frozen)        &  $0.44^{+0.07} _{-0.06}$ &   $0.39$ (frozen)  \\
\multicolumn{4}{c}{----- APEC 1 (fixed to the lowest state param.)-------}   \\
kT$_{\rm APEC1}$   (keV)                  &  $0.24$              &    $0.24$                &    $0.24$       \\
norm$_{\rm APEC1}$ ($10^{-5}$ cm$^{-5}$)  &  $18$                &    $18$                  &    $18$         \\
\multicolumn{4}{c}{----- APEC 2 (fixed to the lowest state param.)-------}   \\
kT$_{\rm APEC2}$   (keV)                   &  $0.76$             &    $0.76$                &    $0.76$   \\
norm$_{\rm APEC2}$ ($10^{-5}$ cm$^{-5}$)   &  $3.5$              &    $3.5$                 &     $3.5$   \\
\multicolumn{4}{c}{-----ADDITIONAL ABSORPTION for the POWER LAW-------}    \\
N$_{\rm H}$ (10$^{22}$ cm$^{-2}$)         & $-$                  &         $-$               &  $0.25^{+0.36} _{-0.25}$   \\
\multicolumn{4}{c}{----- POWER LAW-------}   \\
$\Gamma$                            &  $2.08^{+0.19} _{-0.19}$	 &   $2.24^{+0.29} _{-0.25}$   &     $2.36^{+0.44} _{-0.37}$            \\
norm$_{\rm pow}$  		    & $0.185^{+0.037} _{-0.035}$ &  $0.174^{+0.039} _{-0.036}$ &     $0.172^{+0.040} _{-0.037}$	       \\
UF$_{\rm pow}$$^b$   (\flux)        &  $(3.7^{+0.4} _{-0.4})\times10^{-13}$ & $(3.9^{+0.6} _{-0.5})\times10^{-13}$ &   $(4.3^{+1.5} _{-0.8})\times10^{-13}$  \\  
L$_{\rm pow}$$^{a,b}$   (\lum)          &  $2.1\times10^{32}$        &  $2.3\times10^{32}$                                 &     $2.5\times10^{32}$    \\   
\hline
UF$_{\rm total}$$^b$  (\flux)       &  $(6.0^{+0.4} _{-0.4})\times10^{-13}$ &   $(6.2^{+0.6} _{-0.5})\times10^{-13}$ &  $(6.6^{+1.5} _{-0.8})\times10^{-13}$   \\  
L$_{\rm total}$$^{a,b}$ (\lum)          & $3.5\times10^{32}$                      &  $3.6\times10^{32}$                      &     $3.8\times10^{32}$    \\   
\hline
UF$_{\rm pow}$/ UF$_{\rm total}$    &   61\%                                   &         63\%                              &  64\%  \\
\hline
$\chi^{2}_{\nu}$/dof                &   1.362/46                               &   1.338/45                                &  1.343/45     	 \\
\hline
\hline
\end{tabular}
\footnotesize{\\
$^a$ A source distance of 2.2 kpc is assumed \\
$^b$ All fluxes and luminosities are in the energy range 0.5-10 keV and are corrected for the absorption \\
$^c$ Power law photon index. \\ 
}
\end{center}
\end{table*}
%%%%%%%%%% -------------------------------------------------

	%%%%%%%%%%%%%%%%%%%%%%%%%%%%%%%%%%%%%%%%%%%%
	\subsection{The faint X-ray flare  \label{sec:flarespec}}
	%%%%%%%%%%%%%%%%%%%%%%%%%%%%%%%%%%%%%%%%%%%%

Adopting the same procedure used in the spectral analysis of the intermediate state, we fitted the 
spectra extracted from the faint flare (interval 3 in Fig.~\ref{fig:lc4intselspec})
with a continuum composed of two thermal plasma models
together with a power law. 
The spectral results are listed in Table~\ref{tab:flarespec}. 
The meaning of the three models is the same as outlined in the previous section (Sect.~\ref{sec:interspec}).
The spectrum fitted with one of these models is shown in Fig.\,\ref{fig:bestfit_lowest}.

During the faint flare, the contributed flux from the power law component ia almost 90\% of the total
 unabsorbed 0.5-10 keV flux. 
The power law appears harder with the increasing X-ray intensity (from the lowest state to the faint flare).
The total X-ray luminosity  reaches $10^{33}$ \lum\ in this state.

%%%------------------------------------------------------------------
\begin{table*}
  \caption{Spectroscopy of the emission from the faint flare (\epic\ \pn, \mosuno\ and \mosdue; Sect.~\ref{sec:flarespec}).
}
\label{tab:flarespec}
\vspace{0.0 cm}
\begin{center}
\begin{tabular}{lccc} \hline
 \hline
%\hline
\noalign {\smallskip}
Param.                                  &  Model A               &          Model B         &  Model C          \\
\hline
\noalign {\smallskip}
N$_{\rm H}$ (10$^{22}$ cm$^{-2}$)       & $0.39$ (frozen)        &  $0.57^{+0.28} _{-0.17}$ &   $0.39$ (frozen)  \\
\multicolumn{4}{c}{----- APEC 1 (fixed to the lowest state param.)-------}   \\
kT$_{\rm APEC1}$   (keV)                  &  $0.24$              &    $0.24$                &    $0.24$       \\
norm$_{\rm APEC1}$ ($10^{-5}$ cm$^{-5}$)  &  $18$                &    $18$                  &    $18$         \\
\multicolumn{4}{c}{----- APEC 2 (fixed to the lowest state param.)-------}   \\
kT$_{\rm APEC2}$   (keV)                   &  $0.76$             &    $0.76$                &    $0.76$   \\
norm$_{\rm APEC2}$ ($10^{-5}$ cm$^{-5}$)   &  $3.5$              &    $3.5$                 &     $3.5$   \\
\multicolumn{4}{c}{-----ADDITIONAL ABSORPTION for the POWER LAW-------}    \\
N$_{\rm H}$ (10$^{22}$ cm$^{-2}$)         & $-$                  &         $-$               &  $0.85^{+1.16} _{-0.70}$   \\
\multicolumn{4}{c}{----- POWER LAW-------}   \\
$\Gamma$                            &  $0.97^{+0.24} _{-0.25}$	 &   $1.21^{+0.38} _{-0.34}$   &     $1.61^{+0.72} _{-0.57}$            \\
norm$_{\rm pow}$  		    &  $1.57^{+0.41} _{-0.36}$   &   $1.43^{+0.41} _{-0.35}$   &     $1.31^{+0.41} _{-0.32}$	       \\
UF$_{\rm pow}$$^b$   (\flux)        &  $(1.86^{+0.40} _{-0.35})\times10^{-12}$ & $(1.81^{+0.36} _{-0.32})\times10^{-12}$ &   $(1.93^{+0.83} _{-0.35})\times10^{-12}$  \\  
L$_{\rm pow}$$^{a,b}$   (\lum)          &  $1.08\times10^{33}$        &  $1.05\times10^{33}$                                 &     $1.11\times10^{33}$    \\   
\hline
UF$_{\rm total}$$^b$  (\flux)       &  $(2.09^{+0.40} _{-0.35})\times10^{-12}$ &   $(2.04^{+0.36} _{-0.32})\times10^{-12}$ &  $(2.17^{+0.83} _{-0.35})\times10^{-12}$   \\  
L$_{\rm total}$$^{a,b}$ (\lum)          & $1.21\times10^{33}$                      &  $1.18\times10^{33}$                      &     $1.26\times10^{33}$    \\  
\hline
UF$_{\rm pow}$/ UF$_{\rm total}$    &   89\%                                   &         89\%                              &  89\%  \\
\hline
$\chi^{2}_{\nu}$/dof                &   1.510/11                               &   1.327/10                                &  1.208/10     	 \\
\hline
\hline
\end{tabular}
\footnotesize{\\
$^a$ A source distance of 2.2 kpc is assumed \\
$^b$ All fluxes and luminosities are in the energy range 0.5-10 keV and are corrected for the absorption \\
$^c$ Power law photon index. \\ 
}
\end{center}
\end{table*}
%%%%%%%%%% -------------------------------------------------

%%%%%%%%%%%%%%%%%%%%%%%%%%%%%%%%%%%%%%%%%%%%%%%%%%%%%%%%%%%%%%%%%
\section{Comparison of the lowest state with previous X-ray observations} \label{Sect:comp}
%%%%%%%%%%%%%%%%%%%%%%%%%%%%%%%%%%%%%%%%%%%%%%%%%%%%%%%%%%%%%%%%%

\src\ was previously observed in a low state by \xmm\ in 2007 \citep{Bozzo2010} and by \suz\ in 2009 
\citep{Sidoli2010igr08408}. These observations were performed at different orbital phases 
($\phi$=0.66-0.71 and $\phi$=0.82-0.97, respectively) and with a higher average X-ray flux than observed in 2020.

In previous sections, we have demonstrated that a good deconvolution of the continuum spectrum, also motivated by the known physical properties of the  source, is composed by two thermal plasma models plus a power law.
Since the two thermal models are compatible with X-ray emission from the wind of the supergiant donor,
they should be present also in previous \xmm\ and \suz\ observation with the same parameters 
(assuming no long term variability from the stellar wind in LM Vel). 
We can ascribe the (variable) power law component to emission from the compact object. 
To investigate the hardness and the intensity of the power law emission during the lowest state
in previous observations, we have selected the lowest states emission also in 2007 and in 2009. 

\begin{figure*}
\begin{center}
  \includegraphics[width=6.1cm,angle=-90]{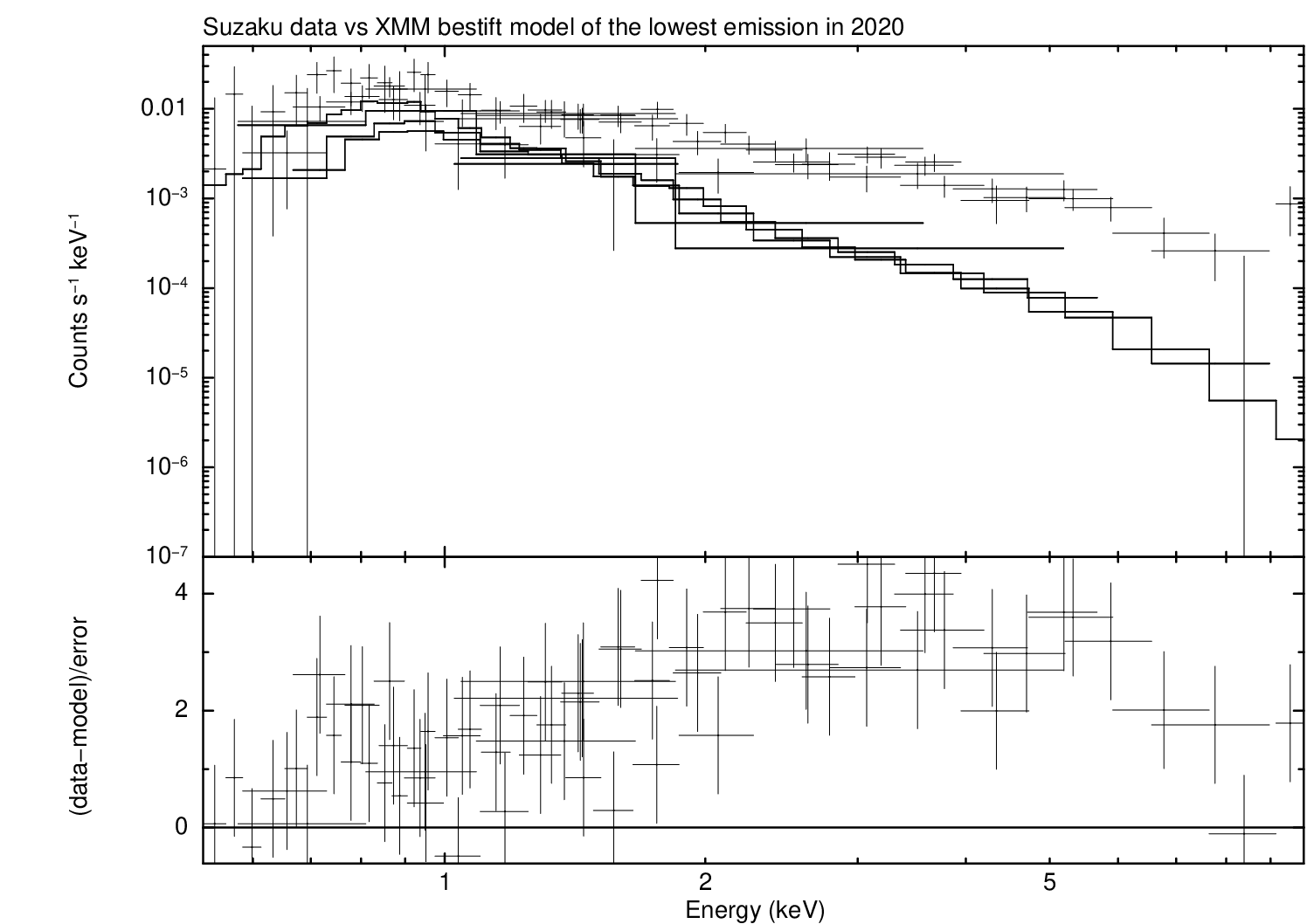}
  \includegraphics[width=6.1cm,angle=-90]{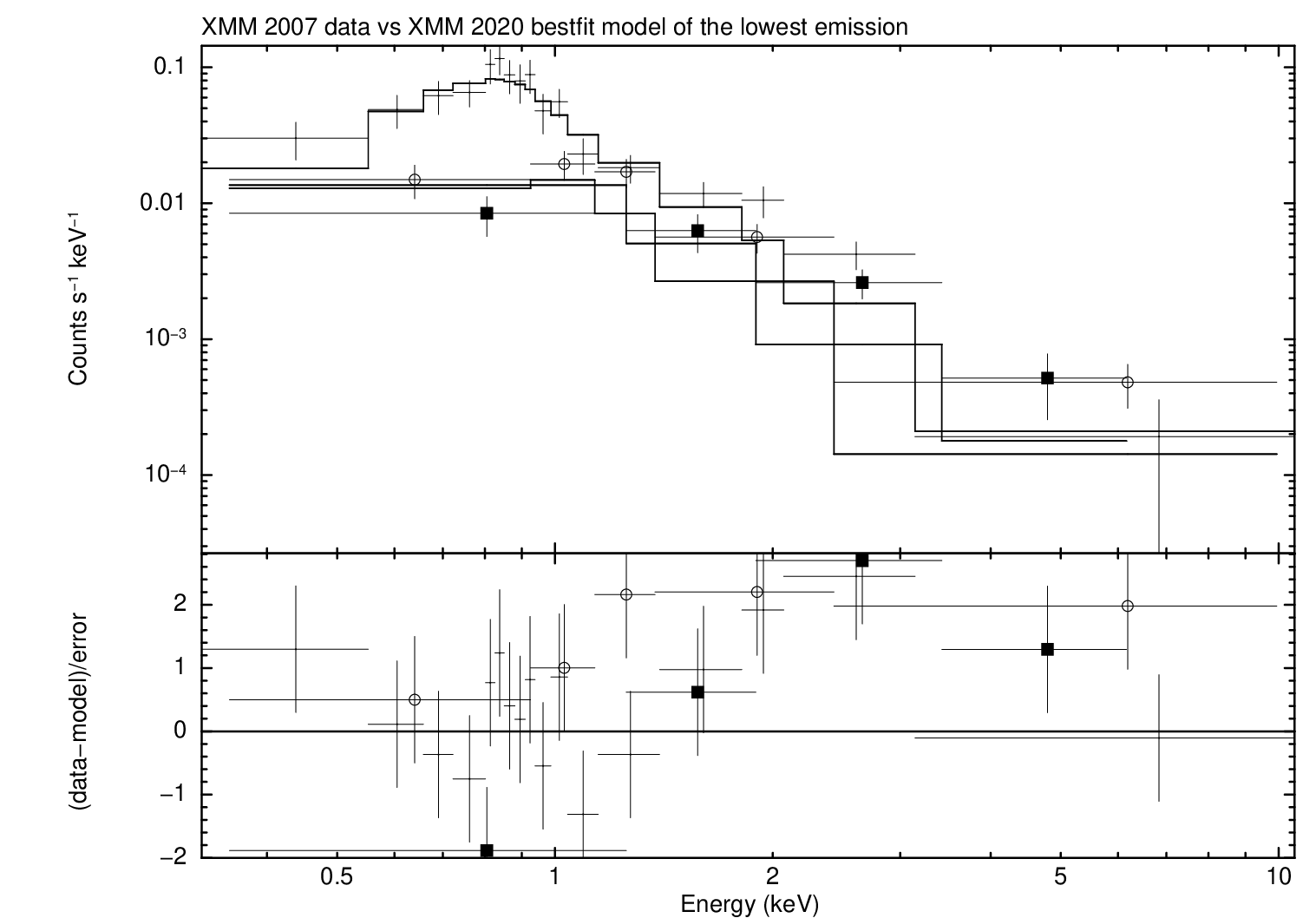}  
  \caption{Comparison of \suz\ (on the left) and \xmm\ observations performed in 2007 (on the right)
with the best-fit of the lowest luminosity state observed in 2020 with \xmm. Residuals are reported in the lower panels
in units of standard deviations.  The meaning of the symbols in the right panel is the following: crosses, empty circles and solid squares mark the EPIC pn, MOS1 and MOS2 spectra, respectively. Clearly, past observations caught \src\ in harder states.
}
\label{fig:comp}
\end{center}
\end{figure*}

In particular, we have applied to the 2007 \xmm\ observation the same intensity filter for the lowest state of the 2020 observation,
i.e. \epic\ \pn\ source count rate lower than 0.05 count s$^{-1}$ 
(the reader can refer to \citealt{Sidoli2010igr08408} and \citealt{Bozzo2010} 
for more details on this \xmm\ observation).
The two \mos\ spectra were extracted using the same good time intervals, obtaining the 2007 \xmm\ spectra reported in Fig.\,\ref{fig:comp} (right panel).
There, we show the residuals with respect to the best fit of the lowest intensity observed in 2020:
while the soft part is well accounted for by the two thermal models (confirming it is a steady spectral component), 
the harder region of the spectrum shows an excess with respect to the power law model fitting the 2020 X-ray emission. 
When let free to vary, the best fit power law component  of the 2007 \xmm\ observation
shows a photon index $\Gamma$=$2.00^{+0.32} _{-0.30}$ 
and an unabsorbed power law flux UF$_{\rm pow}$=(1.34$\pm{0.26})\times10^{-13}$ \flux\ (0.5-10 keV; reduced $\chi^2$=1.064 for 62 degrees of freedom, dof).

We next considered the \suz/XIS observation (0.5-10 keV) performed in December 2009 \citep{Sidoli2010igr08408}, where the source was observed around periastron.
Here we have re-analysed the X-ray spectrum extracted from
the initial part of the 2009
observation, where the source was caught at a low count rate, 
i.e. the persistent spectrum reported in Table\,2 of \citet{Sidoli2010igr08408}. 
In Fig.\,\ref{fig:comp} (left panel) we plot the \suz\ spectrum against the best fit to the lowest intensity observed 
by \xmm\ in 2020. 
We note that in this plot, no cross-calibration constant 
factors have been assumed between \epic\ and XIS spectra (the best fit assumed the \epic\ \pn\ response matrix).
Interestingly, the softest part of the \suz/XIS spectrum is well accounted for by the assumed model, while above 1-2 keV, positive residuals appear above the best fit to the lowest state observed in 2020.
If we leave the power law component to vary freely during the fit, we obtain a best fit (reduced $\chi^2$=1.013 for 71 dof)
photon index $\Gamma$=$1.64\pm{0.16}$ and 
an unabsorbed power law flux UF$_{\rm pow}$=(5.74$^{+0.41} _{-0.49})\times10^{-13}$ \flux\ (0.5-10 keV).

\begin{figure}
\begin{center}
\includegraphics[width=6.0cm,angle=-90]{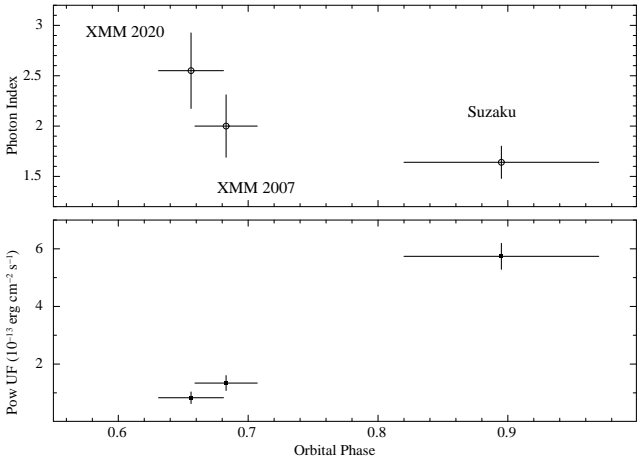}
     \caption{Orbital phase dependence of the power law spectral parameters of lowest intensity states 
observed during the two \xmm\ observations (performed in 2007 and 2020) and the \suz\ one. 
The power law photon index and unabsorbed flux (in units of 10$^{-13}$ \flux) 
are reported in the upper and lower panels, respectively. 
See Sect.\,\ref{Sect:comp} for details.
}
\label{fig:poworb}
\end{center}
\end{figure}

The power law parameters measured during the lowest intensity states in 2007, 2009 and 2020
are reported in  Fig.\,\ref{fig:poworb}, against the orbital phase coverage 
of the correspondent observations (orbital phase $\phi$=1 indicates the periastron).
Harder and brighter power law emission is observed approaching the periastron passage.
The orbital phases have been derived assuming the ephemerides reported by \citet{Gamen2015}. 
We note that extrapolating the uncertainty on the orbital period 
to the epoch of the three observations, the uncertainty on the orbital phase is always $\Delta\phi\lsim$0.01.

%%%%%%%%%%%%%%%%%%%%%%%%%%%%%%%%%%%%%%%%%%%%%%%%%%%%%% 
%%%%%%%%%%%%%%%%%%%%%%%%%%%%%%%%%%%%%%%%%%%%%%%%%%%%%% 

%%%%%%%%%%%%%%%%%%%%%%%%%%%%%%%%%%%%%%%%%%%%%%%%%%%%%
%%%%%%%%%%%%%%%%%%%%%%%%%%%%%%%%%%%%%%%%%%%%%%%%%%%%%% 

%%%%%%%%%%%%%%%%%%%%%%%%%%%%%%%%%%%%%%%%%%%%%%%%%%%%%%%%%%%%%%%%%
\section{Discussion}
%%%%%%%%%%%%%%%%%%%%%%%%%%%%%%%%%%%%%%%%%%%%%%%%%%%%%%%%%%%%%%%%%
 
We have investigated the X-ray properties shown by the SFXT \src\ during an observation
performed  by \xmm\ in June 2020, which caught the source just after the apastron, in its lowest X-ray emission state, to date. 

The source X-ray light curve shows a faint, short flare (which is usual in SFXTs even at low luminosity states, e.g. \citealt{Sidoli2019extras}), 
with emission becoming harder when brighter. 
This behavior suggested to investigate X-ray spectra extracted from three different intervals of source count rate, separately.
The lowest state spectrum is well deconvolved by a three component model, with two-temperature hot plasma model, plus a power law. 
The temperatures of the thermal components (kT=0.24\,keV and kT=0.76\,keV) 
are consistent with the usual X-ray emission observed from O-type supergiants, pointing out to the fact that the donor is a quite normal star.
We ascribe the faint power law component to residual accretion onto the compact object.

The source has been previously caught in a low X-ray state 
(10$^{32}-10^{33}$ erg s$^{-1}$), with a low absorption and evidence for a soft component  \citep{Leyder2007, Bozzo2010, Sidoli2010igr08408}. 
Although these authors have already suggested that the soft component could be due to shocks in the supergiant wind, nevertheless, this is the first time that its presence is clearly established in the source spectrum, thanks to the very low accretion onto the compact object and the high throughput \xmm\ observation.
 
Therefore, the scenario emerging from the spectroscopy of the lowest luminosity state reveals that the unabsorbed flux contributed by the power law component alone (e.g. accretion onto the compact object) is
(8.3$^{+2.4} _{-2.0}$)$\times10^{-14}$ \flux (0.5-10 keV), 
implying a luminosity of only (4.8$\pm{1.4}$)$\times10^{31}$ \lum\ due 
to accretion. In the following sections (\ref{Sect:physics} and \ref{Sect:applied}) we will outline a physical scenario
to explain this very low level of accretion.

Lastly, this picture is confirmed by the comparison with previous low states observed in \src\  with \xmm\ and \suz\ (Sect.\,\ref{Sect:comp}): our best fit to the X-ray emission from the supergiant donor is able to account for the softest region of the spectrum in those observations as well, while a brighter and harder power law component (due to accretion) emerges, towards periastron passage.

 \subsection{Physical picture of the low-luminosity state of SFXTs} \label{Sect:physics}
 
 We assume that accretion in SFXTs proceeds quasi-spherically onto a slowly rotating magnetized NS from the optical companion's stellar wind (\citealt{2014MNRAS.442.2325S}; see, e.g., \citealt{2019NewAR..8601546K} for a recent review). The accreting plasma enters the magnetosphere via the Rayleigh-Taylor instability (RTI). A steady mass accretion rate is possible if plasma arriving at the magnetosphere cools down below some critical temperature \citep{1977ApJ...215..897E}. The Compton cooling by X-rays generated near the NS surface is effective above the  critical X-ray luminosity $L^\dag\simeq 4\times 10^{36}\mathrm{\,erg\,s^{-1}}$. At lower luminosities, a hot, convective atmosphere grows above the NS magnetosphere filling the space up to roughly the Bondi capture radius $R_\mathrm{B}$. 
 
 Through this hot, likely convective shell, a quasi-steady subsonic, settling accretion occurs  at a rate controlled by the plasma cooling mechanism, $\dot M\approx f(u) \dot M_\mathrm{B}$, where 
 $\dot M_\mathrm{B}=4\pi (\rho v_\mathrm{ff}) R_\mathrm{m}^2$ is the Bondi (supersonic) accretion rate (the subscript $m$ means that the plasma density $\rho$ and free-fall velocity are evaluated near the magnetospheric radius $R_\mathrm{m}$). The factor $f(u)\approx (t_\mathrm{ff}/t_\mathrm{cool})^{1/3}<1$ is determined by the ratio of the free-fall time $t_\mathrm{ff}$ to the plasma cooling time $t_\mathrm{cool}$. At low X-ray luminosities, $L_\mathrm{X}\ll L^\dag$, the quasi-steady plasma entry rate is mediated by radiative cooling. In this case \citep{2013MNRAS.428..670S}
 \begin{equation}
 \label{e:furad}
     f(u)_\mathrm{rad}\simeq 0.1 L_{36}^{2/9}\mu_{30}^{2/27}\,.
 \end{equation}
 Here $L_{36}\equiv L_x/(10^{36}\mathrm{erg\, s^{-1}})=0.1\times10^{16}[\mathrm{g\,s^{-1}}]\dot  M_\mathrm{x,16}c^2$ is the accretion X-ray luminosity, the NS magnetic moment is in units of
$10^{30}~\mathrm{G\,cm^3}$, and the numerical coefficient is calculated for a 1.5 $M_\odot$ NS.
 It is convenient to express this factor through the gravitational capture Bondi rate $\dot M_\mathrm{B}$, which is related to 
 the accretion rate for the radiative cooling as $\dot M_\mathrm{x,16}\approx 0.05\times \dot M_\mathrm{B,16}^{9/7}\mu_{30}^{2/21}$ \citep{2019MNRAS.485..851Y}:
 \begin{equation}
     \label{e:furadB}
      f(u)_\mathrm{rad}\simeq 0.05 \dot M_\mathrm{B,16}^{2/7}\mu_{30}^{2/21}\,.
 \end{equation}
We stress that equation (\ref{e:furad}) is appropriate when the accretion X-ray luminosity is known, while Eq. (\ref{e:furadB}) uses the Bondi mass capture rate that is not directly measurable but can be estimated from stellar wind and the binary system's parameters.
 In this model, sporadic X-ray outbursts in SFXTs can be due to magnetospheric instability triggered, for example, by magnetic reconnection with external stellar wind magnetic field carried by plasma blobs close to the magnetospheric boundary (\citealt{2014MNRAS.442.2325S} and below).
 
The condition for RTI to be effective can be formulated as the requirement for plasma to cool down below a critical temperature depending on the magnetosphere's curvature \citep{1977ApJ...215..897E}. In a convective shell around the NS magnetosphere, the convection lifts up the hot gas from the magnetospheric boundary. Therefore, for an effective RTI to occur, the 
plasma cooling time near the magnetosphere should be shorter than the convective overturn time, $t_\mathrm{cool}<t_\mathrm{conv}$. The radiative cooling time is \citep{2013MNRAS.428..670S} 
\begin{equation}
    \label{e:trad}
    t_\mathrm{rad}\approx 300 \mathrm{\,[s]}\mu_{30}^{2/3}\dot M_\mathrm{x,16}^{-1}\approx 6000 \mathrm{\,[s]}\mu_{30}^{4/7}\dot M_\mathrm{B,16}^{-9/7}
\end{equation}
The longest convection overturn time is commensurable with the free-fall time from the Bondi radius, $R_\mathrm{B}=2GM/v_\mathrm{w}^2$: 
\begin{equation}
    t_\mathrm{conv}=\zeta_c t_\mathrm{ff}(R_\mathrm{B})\simeq  400\mathrm{\,[s]}\zeta_c v_8^{-3}\,,
\end{equation}
where the dimensionless factor $\zeta_c\gtrsim 1$, the stellar wind velocity relative to the NS $v_\mathrm{w}=10^8\mathrm{cm\,s^{-1}}v_8$. Therefore, for radiative plasma cooling, we expect the RTI to operate when
\begin{equation}
\label{e:RTI}
    \dot M_\mathrm{B,16}>8.2\zeta_c^{-7/9}\mu_{30}^{4/9}v_8^{7/3}\,,
\end{equation}
For a NS moving around an  optical star with a mass-loss rate of $\dot M_O=10^{-6}(M_\odot/\mathrm{yr})\dot M_{O,-6}$, the Bondi capture rate is $\dot M_\mathrm{B,16}\approx 10^4/(2\pi) \dot M_{O,-6}(R_\mathrm{B}/r)^2$ (here $r$ is the distance from the optical star to the NS). Thus, the condition (\ref{e:RTI}) can be recast to the form
\begin{equation}
    \label{e:RTI1}
    \frac{r}{R_\odot}\lesssim 8 v_8^{-19/6}\dot M_{O,-6}^{1/2} \zeta_c^{7/18}\mu_{30}^{-2/9}\,.
\end{equation}
For example, it is easy to check that the condition (\ref{e:RTI1}) can be met for systems like Vela X-1 with $v_8\sim 0.7$, $\mu_{30}=1.2$ and $r\approx 50 R_\odot$ for $\zeta_c\sim$ a few\footnote{In Vela X-1, however, the Compton plasma cooling should be more effective most of the time; see \cite{2012MNRAS.420..216S, 2013MNRAS.428..670S}.}.

What happens if the accretion rate onto NS in a binary system drops below the critical value (\ref{e:RTI}), and the magnetosphere turns out to be Rayleigh-Taylor stable? This could be the case for a NS moving in an elliptic orbit near apastron. Then, the quasi-static plasma entry can be sustained by
(i) the diffusion through the magnetosphere, (ii) the magnetospheric cusp instability, or (iii) magnetic reconnection with magnetic field carried out by accreting stellar wind blobs \citep{1984ApJ...278..326E}.

Of these possible physical mechanisms, the most effective is the entry via turbulent plasma diffusion in the Bohm regime. An extreme upper bound on the plasma entry rate can be obtained from Eq. (59) of \cite{1984ApJ...278..326E} applied to the settling accretion regime.  
In this case, we should take into account that in the settling accretion regime, the magnetospheric radius differs from the reference Alfv\'en value 
$R_A=\left(\frac{\mu^2}{\dot M \sqrt{GM}}\right)^{2/7}$ (see \citealt{2012MNRAS.420..216S}): 
\begin{equation}
R_\mathrm{m}\approx \left[\frac{4\gamma}{(\gamma-1)}K_2 f(u)\right]^{2/7}R_A\,.
\end{equation} 
For accretion of a monoatomic gas with adiabatic index $\gamma=5/3$, the factor in the brackets in the above equation takes into account the
magnetospheric currents boundary screening \citep{1976ApJ...207..914A} and is equal to 76.2. For the radiative plasma cooling, the magnetospheric radius reads \citep{2013MNRAS.428..670S}
\begin{equation}
\label{e:Rm}
    R_\mathrm{m}\approx 10^9 [\mathrm{cm}]\mu_{30}^{16/27} L_\mathrm{36}^{-2/9}\simeq 2\times 10^9[\mathrm{cm}]\mu_{30}^{4/7}\dot M_\mathrm{B,16}^{-2/7}\,.
\end{equation}
(In the second equation we have changed the X-ray luminosity $L_{36}$ by $\dot M_\mathrm{B,16}$ in a way as
in Eqs. (\ref{e:furad}) and (\ref{e:furadB})).
As the Bondi mass accretion rate enters Eq. [32] of \cite{1984ApJ...278..326E}, determining the Bohm diffusion through magnetosphere via the Alfven radius $R_A$, we should change $\dot M \to \dot M_B/(76 f(u)_\mathrm{rad})$ 
in Eq. [59] of \cite{1984ApJ...278..326E}. Then, using formula (\ref{e:furadB}) for $f(u)_\mathrm{rad}$ through $\dot M_B$, we arrive at a maximum suppression factor of the free-fall Bondi accretion rate due to the Bohm-diffusion  in the radiation cooling settling regime:
\begin{equation}
\label{e:dotMdif}
   \frac{\dot M_\mathrm{dif}}{\dot M_\mathrm{B}}\simeq 3.3 \times 10^{-5} 
   \dot M_\mathrm{B,16}^{-15/98}\mu_{30}^{-5/98}\,.
\end{equation}
Note a very weak dependence on the (unknown) NS magnetic field.
For practical use, we can recast formula (\ref{e:dotMdif}) to the form
\begin{equation}
    \label{e:dMdif}
    \dot M_\mathrm{dif}\simeq 3.3 \times 10^{11} [\mathrm{g\,s^{-1}}] 
   \dot M_\mathrm{B,16}^{83/98}\mu_{30}^{-5/98}\,.
\end{equation}
Thus, in SFXTs, at a given Bondi gravitational capture rate $\dot M_\mathrm{B}$ determined by the orbital parameters and stellar wind properties of the optical star, there should be a minimum possible accretion X-ray luminosity in the low states of SFXTs, 
\begin{equation}
\label{e:Lxmin}
L_\mathrm{X,min}\sim 3.3\times 10^{31}[\mathrm{erg\,s^{-1}}]\dot M_\mathrm{B,16}^{83/98}\mu_{30}^{-5/98}\,.
\end{equation}

%%%%%%%%%% ---------------------------------------------------------
\subsection{Application to \src} \label{Sect:applied}  

\textbf{A) Minimum X-ray luminosity}.
 Eqs. \ref{e:dotMdif} and \ref{e:Lxmin} suggest that there should exist
 an almost universal lower limit to mass accretion rate in SFXTs with the radiation-cooling settling accretion. Its value is proportional to the Bondi mass accretion rate, $\dot M_\mathrm{dif}\propto \dot M_\mathrm{B}^{83/98}$. By knowing (or assuming) the stellar wind mass-loss rate, the Bondi capture rate can be estimated from the binary system parameters. For example, we can use the recent semi-analytical analysis of \cite{2021MNRAS.501.2403B} showing that for  \src, the expected Bondi mass accretion rate should be around $3\times 10^{15}-10^{16}  ~\mathrm{g\,s^{-1}}$ (see their Fig. 7, left panel).  
  With these Bondi capture rates, the condition (\ref{e:RTI}) can be violated 
  at the orbital phases of the \xmm\  observations, which turn out RTI-stable.
  Substituting $\dot M_\mathrm{B,16} \sim 1$ into Eq. \ref{e:Lxmin}, we get 
 for \src\, $L_\mathrm{X,min}\approx 3.3\times 10^{31}~\mathrm{erg\,s^{-1}}$, slightly below the observed lowest X-ray luminosity.
 The factor of two uncertainty in the stellar wind parameters of LM Vel are possible \citep{Hainich2020}. For example, a slight increase in the optical star mass-loss rate, or an insignificant decrease in the stellar wind velocity, could easily bring the expected residual X-ray luminosity of \src\ in agreement with the observed $5\times 10^{31}~\mathrm{erg\, s^{-1}}$.

We note that \cite{2021MNRAS.501.2403B} have introduced by hand an accretion suppression factor $\chi=7\times 10^{-5}$ to match the observed low luminosities of \src\, 
 at periastron (see Fig 7, right panel in \citealt{2021MNRAS.501.2403B}). In our formulation, the low X-ray luminosity between SFXT flares is due to plasma entry into magnetosphere at the radiation-driven settling accretion stage (see \citealt{Sidoli2019extras} for more details).
 Therefore, this factor naturally arises from our physical model (see Eq. \ref{e:dotMdif}). No additional 'suppression' of accretion is needed.

Thus, our estimates show that the expected maximum diffusion mass accretion rate in the low states of SFXTs can be as low as a few $ 10^{31}~\mathrm{erg\,s^{-1}}$, in agreement with what is observed.
 
 \vskip\baselineskip
 \noindent\textbf{B) The lack of strong flaring activity.}
One may wonder why we do not see bright X-ray flares at apastron phases of \src. Note, in the first place, that the orbital phases around apastron of \src\ are RTI-stable (the condition (\ref{e:RTI}) is violated for the expected Bondi capture rates). In our formulation, the lack of bright outbursts means that the possible magnetic reconnection is less effective at orbital phases $\sim 0.1-0.6$ around apastron. 
To see this, consider the magnetic reconnection of a plasma blob\footnote{The magnetized blobs near the NS magnetosphere are different from the stellar-wind 'clumps'; they can appear in the convective settling shell even in the case of gravitational capture of an almost homogeneous stellar wind.} with size $\lambda R_\mathrm{m}$ ($\lambda\ll 1$), density $\rho'\sim \rho_\mathrm{m}$ and magnetic field $B'=\alpha B_\mathrm{m}$ with the  magnetospheric field $B_\mathrm{m}$. The reconnection time is 
$t_\mathrm{r}=\lambda R_\mathrm{m}/u_\mathrm{r}$, the reconnection rate is $u_\mathrm{r}=\epsilon_\mathrm{r} u_A$ where $u_A=B'/\sqrt{4\pi \rho'}$ is the  Alfvenic velocity in the weaker-field blob, and $\epsilon_\mathrm{r}\sim 0.01-0.1$ is the reconnection efficiency. By definition, the magnetospheric radius at the settling accretion stage is $B_\mathrm{m}^2/4\pi\sim \rho_\mathrm{m} c_s^2$, where $c_s^2=(2/5)GM/R_\mathrm{m}$ is the thermal sound velocity near the magnetosphere \citep{2012MNRAS.420..216S}. Noticing that the free-fall velocity at the magnetosphere is $u_\mathrm{ff}^2=2GM/R_\mathrm{m}$, we find: $t_\mathrm{r}\sim (\lambda \sqrt{5}/\alpha\epsilon_\mathrm{r}) t_\mathrm{ff}(R_\mathrm{m})$. 

For an effective reconnection to occur, the reconnection time should be shorter than the time the blob spends near the magnetospheric boundary, the convection overturn time, 
$t_\mathrm{r}<t_\mathrm{ff}(R_\mathrm{B})
=\zeta_c t_\mathrm{ff}(R_\mathrm{B})$.
 
\begin{equation}
%\begin{multline}
\label{e:trtconv}
  \frac{t_\mathrm{r}}{\zeta_c t_\mathrm{ff}(R_\mathrm{B})}=\frac{\lambda\sqrt{5}}{\zeta_c\alpha\epsilon_\mathrm{r}}\left(\frac{R_\mathrm{m}}{R_\mathrm{B}}\right)^{3/2}\approx 
  \frac{\lambda}{\alpha\zeta_c}\frac{0.03}{\epsilon_\mathrm{r}}\dot M_\mathrm{B,16}^{-1/3}\mu_{30}^{6/7}v_8^3  \lesssim 1\,.
%\end{multline}
\end{equation}
It is seen that this ratio strongly depends on the wind velocity. 
The numerical coefficient here is determined by the scale $\lambda$ and amplitude $\alpha$ of the magnetic cell. During turbulent plasma infall, $B_t'^{2}\lesssim 4\pi \rho_m c_s^2$, $\alpha\lesssim 1$, and the turbulent cell size can be $\lambda_t\lesssim 1$.
In the case of \src, the relative wind velocity increases from $\sim 500$~km~s$^{-1}$ at periastron to $\sim 1400$~km~s$^{-1}$ at apastron \citep{Hainich2020} changing the ratio $t_\mathrm{r}/t_\mathrm{ff}$ by the factor $\sim 30$ over the orbit. 
Thus, the violation of the inequality (\ref{e:trtconv}) can explain the transition from the flaring state near periastron to quiescent behavior close to apastron.

\vskip\baselineskip
\noindent
\textbf{C) Magnetized stellar wind effects in SFXTs.}
The effect of a  magnetized stellar wind in HMXBs can be twofold. First, the magnetic reconnection of blobs with embedded magnetic field $B'$ comparable to the field at the magnetospheric boundary $B_\mathrm{m}$ can dramatically disturb (even open) the NS magnetosphere leading to bright SFXT outbursts \citep{2014MNRAS.442.2325S}. Second, the magnetic reconnection of small-size blobs at the base of the shell around the magnetosphere, which does not strongly disturb  the magnetospheric boundary, would additionally heat the plasma.
This heating could further hinder the RTI development, hamper the plasma entry, and strengthen the convection in the shell \citep{2012MNRAS.420..216S}. It is tempting to suggest that the additional reconnection-induced plasma heating in magnetized plasma blobs occurs even at the periastron of \src\ (where the RTI condition (\ref{e:RTI}) is evidently met). This could be responsible for an increased SFXT activity at the periastron  but not full turning-on of the RTI-mediated accretion on the NS in this source.

The moderate flare of \src\, detected during our observations (see Fig.~\ref{fig:2lc_ratio}) can be 
the manifestation of a sporadic magnetic reconnection in 
the RTI-stable hot shell.
Indeed, the mass of the settling shell is $\Delta M_\mathrm{rad}\simeq 3.7\times 10^{15}(v_\mathrm{w}/1000\mathrm{\,km\,s^{-1}})^{-3}$~g \citep{2014MNRAS.442.2325S}. If the magnetospheric instability is sporadically caused by the magnetic reconnection with a  large magnetized stellar wind clump with $\lambda\sim 1$, the characteristic time of an outburst will be of order of the free-fall time from the Bondi radius $t_\mathrm{ff}\sim 10^3$~s (see Fig. 2). Therefore, the X-ray luminosity of the outburst should be about $10^{33}\,\mathrm{erg\,s^{-1}}$, close to the  observed value.

Among O-type stars, the stars with Of?p spectral types are recognized as  strongly magnetic \citep{Walborn2010, Grunhut2017}. In these stars, as well as 
as in some magnetic O-dwarfs, magnetic field strongly influences the dynamics of stellar winds. In X-rays, the magnetic O-type dwarfs are harder and brighter compared to their non-magnetic counterparts  \citep[e.g.][]{Schulz2000,Shenar2017}.
Direct searches for a regular magnetic field in  LM Vel have failed so far \citep{2018MNRAS.474L..27H}. However, a weak, 50--100\,G, magnetic fields with a complex topology have been detected in the spectroscopically similar O9.7Ib supergiant $\zeta$~Ori \citep{Bouret2008}. The X-ray properties of $\zeta$~Ori are usual for its spectral type \citep{Waldron2007} with no signs of a power law continuum detected in the X-ray spectrum of this 
well studied star.

%%%%%%%%%%%%%%%%%%%%%%%%%%%%%%%%%%%%%%%%%%%%%%%%%%%%%%%%
\section{Conclusions} \label{Sect:concl}
%%%%%%%%%%%%%%%%%%%%%%%%%%%%%%%%%%%%%%%%%%%%%%%%%%%%%%%%
 
 The \xmm\ observation of the SFXT \src\ performed in 2020 caught the source in a very low level of X-ray activity, with a flux of 3.2$\times10^{-13}$ \flux\ (0.5-10 keV, corrected for the interstellar absorption). The X-ray spectrum is well described by a three-component model: two thermal plasma models (showing temperatures of kT$_1$=0.24 keV and kT$_2$=0.76 keV), together with a power law ($\Gamma$=2.55) that dominates emission above $\sim$2 keV and contributes about 25\% of the X-ray flux in the 0.5-10 keV energy range.
 
 We have argued that the power law X-ray component of \src\, observed during \xmm\ observations at the orbital phases $\phi\sim 0.65$ at a level of $L_\mathrm{X}\simeq 5\times 10^{31}\,\mathrm{erg\,s^{-1}}$ can be explained by the diffusion plasma entry rate into the NS magnetosphere at the radiation-dominated settling accretion stage. The mild flaring activity at these orbital phases of \src\, can be due to reconnection in magnetized stellar wind blobs arriving at the magnetosphere, as proposed by us earlier \citep{2014MNRAS.442.2325S,Sidoli2019extras}. Notably, the diffusion accretion rate at low-states of SFXTs (Eq. \ref{e:dotMdif}) is almost independent of the NS magnetic field and almost linearly depends on the Bondi accretion rate from the stellar wind of the optical star.
The minimum possible accretion luminosity in this case would be $L_\mathrm{X,min}\sim 3\times 10^{31}$~erg s$^{-1}$. The timing X-ray properties of the diffusion entry should be different from those supposedly observed in SFXTs where the Rayleigh-Taylor instability operates 
\citep{Sidoli2019extras}.
It would be interesting to further investigate the low (unflared) state of other SFXTs to check our models.

The observations of \src\ at its lowest state allowed us to detect 
the intrinsic X-ray emission from the O-type supergiant donor star. The properties of the donor star X-ray emission are very similar to those of other O-supergiants: $\log{L_{\rm X}}\approx 32.7$\,[erg\,s$^{-1}$] and the X-ray spectrum well described by thermal, collisionally ionized plasma model with $T_{\rm X}\approx 3$\,--\,$10$\,MK. The discovery of average X-ray properties  further highlights that the donor star is not a peculiar object but has usual, for its spectral type, stellar and wind properties.

%%%%%%%%%%%%%%%%%%%%%%%%%%%%%%%%%%%%%%%%%%%%%%%%%%%%%%%%
%%%%%%%%%%%%%%%%%%%%%%%%%%%%%%%%%%%%%%%%%%%%%%%%%%%%%%%%
%%%%%%%%%%%%%%%%%%%%%%%%%%%%%%%%%%%%%%%%%%%%%%%%%%%%%%%%

\begin{appendix}
\section{Lowest emission state: additional models  \label{sec:app}}

For the sake of completeness, we report in Table~\ref{tab:quiesc} the spectral parameters obtained fitting the spectrum of the lowest state with the six models
discussed in Sect.~\ref{sec:quiesc}. All models resulted into equally good deconvolutions of the spectrum. 
However, the expected X-ray emission from the supergiant donor itself, favours the two {\sc apec} models (Model 3, in Table~\ref{tab:quiesc}). 
Nevertheless, some issues remain when adopting Model~3, as discussed in Sect.~\ref{sec:quiesc}.
In conclusion, for better clarity, our preferred, final deconvolution of the spectrum is reported in Table~\ref{tab:quiesc_3apecs} (Model 2).

We note that in Table~\ref{tab:quiesc} the uncertainty on the emission measure of the thermal emission model
{\sc apec} in {\sc xspec} (EM$_{\rm APEC}$) 
has been derived  only from the error on the normalization of the thermal model
(i.e. no uncertainty on the distance has been considered).
We have always adopted solar abundances in {\sc apec} model.

%%%------------------------------------------------------------------
\begin{table*}[ht]
  \caption{Spectroscopy of the  lowest luminosity state (\epic\ \pn, \mosuno\ and \mosdue)  with statistically acceptable models (see Sect.~\ref{sec:app} and Sect.~\ref{sec:quiesc} for details and issues).
}
\label{tab:quiesc}
\vspace{0.0 cm}
\begin{center}
\begin{tabular}{lcccccc} \hline
 \hline
%\hline
\noalign {\smallskip}
Param.                                   &  Model 1 $\,^{(d)}$                 &    Model 2 $\,^{(d)}$                 &  Model 3 $\,^{(d)}$             &   Model 4 $\,^{(d)}$              &    Model 5 $\,^{(d)}$              &    Model 6 $\,^{(d)}$ \\
%-------------------------------------------------------------------------------------------------------------------------------------------------------------------------        
\hline
\noalign {\smallskip}
N$_{\rm H}$ (10$^{22}$ cm$^{-2}$)     & $0.13^{+0.06} _{-0.07}$     & $0.93^{+0.21}_{-0.18}$     & $0.014^{+0.040}_{-0.014}$   & $0.62^{+0.11} _{-0.12}$  & $1.22 ^{+0.26} _{-0.23}$   & $0.30^{+0.14} _{-0.12}$  \\
\multicolumn{7}{c}{---------------------\quad Additional absorber \quad---------------------}   \\
N$_{\rm H\,pcfabs}$ (10$^{22}$ cm$^{-2}$)   & $-$                    & $-$                         & $-$                    & $-$                     & $-$         &    $5.9 ^{+2.5} _{-2.2}$    \\                           
cov.frac$_{\rm pcfabs}$                     & $-$                    & $-$                         & $-$                    & $-$                     & $-$         &    $0.88 ^{+0.05} _{-0.13}$  \\ 
\multicolumn{7}{c}{-----------------\quad Softer thermal component \quad-----------------}   \\
kT$_{\rm BB}$   (keV)                       & $-$                       & $8.5^{+1.0}_{-0.9}\times10^{-2}$  & $-$                   & $-$           &    $7.0 ^{+0.9} _{-0.7}\times10^{-2}$   &  $-$  \\
norm$_{\rm BB}$                             & $-$                       & $3.28^{+9.8}_{-2.6}\times10^{4}$    & $-$                   & $-$           &    $5.9 ^{+44.} _{-5.1}\times10^{5}$     &  $-$  \\
R$_{\rm BB}$$^a$  (km)                      & $-$                       & $40^{+40}_{-20}$                 & $-$                   & $-$           &    $170 ^{+320} _{-110}$                &  $-$  \\
%%--------------
kT$_{\rm APEC}$   (keV)                   &  $0.66^{+0.08} _{-0.06}$  & $-$                     & $0.72^{+0.04}_{-0.07}$    & $0.25^{+0.04} _{-0.02}$ &   $-$     &   $0.64^{+0.09} _{-0.06}$  \\
norm$_{\rm APEC}$ ($10^{-5}$ cm$^{-5}$)   &  $3.8^{+0.8} _{-0.6}$  & $-$                     & $3.2^{+0.5}_{-0.4}$    & $60  ^{+40} _{-30}$     &   $-$     &   $50  ^{+70} _{-30}$     \\
%----
EM$_{\rm APEC}$$^a$ ($10^{54}$ cm$^{-3}$) &  $2.18^{+0.45} _{-0.36}$  & $-$                     & $1.85^{+0.29}_{-0.23}$    & $35  ^{+23} _{-17}$     &   $-$     &   $30  ^{+40} _{-20}$      \\
%-------------------------------------------------------------------------------------------------------------------------------------------------------------------------
UF$_{\rm soft}$$^b$   (\flux)     & $7.21^{+1.37}_{-1.15}\times10^{-14}$ & $2.75^{+3.84}_{-1.47}\times10^{-12}$ & $6.2^{+1.0}_{-0.8}\times10^{-14}$ & $5.9^{+3.4}_{-2.4}\times10^{-13}$   &  $1.0^{+2.1}_{-0.7}\times10^{-11}$  &   $9.9^{+13.4}_{-6.2}\times10^{-13}$  \\  
\multicolumn{7}{c}{-----------------\quad Harder thermal component \quad-----------------}    \\
kT$_{\rm BB}$   (keV)                      &  $-$                 & $0.45^{+0.06}_{-0.05}$         & $-$                   & $0.53^{+0.06} _{-0.06}$                 &    $-$    & $-$ \\
norm$_{\rm BB}$                            &  $-$              & $0.19^{+0.16}_{-0.08}$            & $-$                   & $6.9 ^{+4.3} _{-2.7}\times10^{-2}$      &    $-$    & $-$ \\
R$_{\rm BB}$$^a$  (km)                     &  $-$                      & $0.096^{+0.03}_{-0.02}$    & $-$                   & $5.8 ^{+1.6} _{-1.3}\times10^{-2}$      &    $-$    & $-$ \\
%-------------------------------------------------------------------------------------------------------------------------------------------------------------------------
kT$_{\rm APEC}$   (keV)                    &  $-$                      &  $-$                    & $3.24^{+0.92}_{-0.67}$   &  $-$                     &   $-$         &  $-$ \\
norm$_{\rm APEC}$ ($10^{-5}$ cm$^{-5}$)    &  $-$                      &  $-$                    & $4.55^{+0.72}_{-0.69}$   &  $-$                     &   $-$         &  $-$ \\
%----
EM$_{\rm APEC}$$^a$ ($10^{54}$ cm$^{-3}$)  &  $-$                      &  $-$                    & $2.63^{+0.42}_{-0.40}$   &  $-$                     &   $-$         &  $-$ \\
%-------------------------------------------------------------------------------------------------------------------------------------------------------------------------
UF$_{\rm hard}$$^b$   (\flux)              &  $-$             & $7.7^{+1.5}_{-1.2}\times10^{-14}$ & $6.3^{+1.0}_{-0.5}\times10^{-14}$   & $5.5^{+0.8}_{-0.7}\times10^{-14}$     &    $-$    &   $-$   \\  
%-------------------------------------------------------------------------------------------------------------------------------------------------------------------------
\multicolumn{7}{c}{--------------------\quad   Power law component   \quad--------------------}    \\
$\Gamma$$^c$                            &  $2.54^{+0.33} _{-0.34}$     &  $-$                    &  $-$                  &  $-$                     &   $3.49^{+0.43}_{-0.39}$  & $4.48^{+1.21}_{-0.95}$    \\  
norm$_{\rm pow}$  (10$^{-2}$)             &  $2.87^{+0.82} _{-0.74}$     &  $-$                    &  $-$                  &  $-$                     &   $3.27^{+0.75}_{-0.68}$  & $2.5^{+1.5}_{-1.5}$ \\  
%-------------------------------------------------------------------------------------------------------------------------------------------------------------------------
UF$_{\rm pow}$$^b$   (\flux)       & $8.4^{+1.1}_{-1.6}\times10^{-14}$ &  $-$                    &  $-$                  & $-$             &  $2.8^{+1.5}_{-0.9}\times10^{-13}$  & $7.9^{+18.5}_{-5.4}\times10^{-13}$  \\  
%-------------------------------------------------------------------------------------------------------------------------------------------------------------------------
\hline
%-------------------------------------------------------------------------------------------------------------------------------------------------------------------------
UF$_{\rm total}$$^b$  (\flux)  & $1.56^{+0.19}_{-0.20}\times10^{-13}$   & $2.8^{+3.8}_{-1.5}\times10^{-12}$  & $1.25^{+0.12}_{-0.12}\times10^{-13}$ & $6.5^{+3.4}_{-2.4}\times10^{-13}$   &  $1.03^{+2.13}_{-0.66}\times10^{-11}$  & $1.8^{+3.1}_{-1.2}\times10^{-12}$   \\ 
L$_{\rm total}$$^{a,b}$ (\lum)               & $9.0\times10^{31}$ &   $1.6\times10^{33}$    & $7.2\times10^{31}$    & $3.8\times10^{32}$           & $6.0\times10^{33}$      &  $1.0\times10^{33}$     \\ 
%-------------------------------------------------------------------------------------------------------------------------------------------------------------------------
\hline
UF$_{\rm soft}$/ UF$_{\rm total}$        &   46\%                 &    97\%                 &     50\%              & 91\%                         &     98\%                  & 55\%  \\
\hline
$\chi^{2}_{\nu}$/dof                      &              1.046/143  &    0.938/143             &   1.098/143            &   0.950/143                   &   0.989/143                & 0.954/141  \\  
\hline
\hline
\end{tabular}
\footnotesize{\\
$^a$ A source distance of 2.2 kpc is assumed \\
$^b$ All fluxes and luminosities are in the energy range 0.5-10 keV and are corrected for the absorption \\
  $^c$ Power law photon index. \\
  $\,^{(d)}$ The models, using the {\sc xspec} synthax, are the following:   
{\sc TBabs * (apec + pegpwrl)} (Model 1), 
{\sc TBabs * (bbodyrad + bbodyrad)} (Model 2),
{\sc TBabs * (apec + apec)} (Model 3),     
{\sc TBabs * (apec + bbodyrad)} (Model 4), 
{\sc TBabs * (bbodyrad + pegpwrl)} (Model 5) and    
{\sc TBabs * pcfabs * (apec + pegpwrl)} (Model 6). \\
}
\end{center}
\end{table*}
%%%%%%%%%% -------------------------------------------------

\end{appendix}
%%%%%%%%%%%%%%%%%%%%%%%%%%%%%%%%%%%%%%%%%%%%%%%%%%%%%%%%%%

%----------
\begin{acknowledgements}
This work is based on data from observations with \xmm, an 
ESA science mission with instruments and contributions directly funded by ESA Member States and NASA.
We thank our anonymous referee for the very constructive report and useful comments that helped to improve the clarity of our writing.
LMO  acknowledges  partial support
by the Russian Government Program of Competitive Growth of Kazan
Federal University. The work of KP was partially supported by RFBR grant 19-02-00790.

\end{acknowledgements}

\bibliographystyle{aa}

\end{document}